\documentclass[a4paper,12pt]{article}
\usepackage{amsmath}
\usepackage{amssymb}
\usepackage{amsthm}
\usepackage{latexsym}
\usepackage{mathrsfs}
\usepackage[dvips]{graphicx}
\usepackage[colorlinks=true, citecolor=blue]{hyperref}
\usepackage{pstricks}
\usepackage{color}
\usepackage{tikz}
\usetikzlibrary{decorations.pathreplacing}
\usetikzlibrary{calc}
\usetikzlibrary{shapes.geometric}
\usetikzlibrary{decorations.markings}
\usetikzlibrary{decorations.pathmorphing}
\usepackage{pstricks}
\def\ben{\begin{equation}}
\def\een{\end{equation}}
\def\bena{\begin{eqnarray}}
\def\eena{\end{eqnarray}}
\def\non{\nonumber}

\renewcommand{\ell}{l}

\newcommand{\half}{\frac{1}{2}}
\newcommand{\quater}{\frac{1}{4}}
\newcommand{\mz}{\mathbb{Z}}

\theoremstyle{definition}
\newtheorem{thm}{Theorem}

\newtheorem{lemma}{Lemma}

\newtheorem{conjecture}{Conjecture}

\theoremstyle{definition}

\renewcommand{\H}{\mathscr{H}}
\newcommand{\M}{\mathscr{M}}
\renewcommand{\pounds}{{\mathscr L}}

\newcommand{\D}{\mbox{d}}
\newcommand{\N}{\mathscr{N}}

\newcommand{\y}[1]{\phantom{#1}}

\newcommand{\mr}{\mathbb{R}}
\newcommand{\mc}{\mathbb{C}}
\newcommand{\I}{\mathscr{I}}

\newcommand{\V}{\mathcal{V}}

\newcommand{\T}{\mathbb{T}}
\newcommand{\e}{{\rm e}}

\begin{document}

\title{Black hole uniqueness theorems in higher dimensional spacetimes}

\author{Stefan Hollands$^{1}$\thanks{\tt HollandsS@cardiff.ac.uk}
and Akihiro Ishibashi$^{2}$\thanks{\tt akihiro@phys.kindai.ac.jp}
\\ \\
{\it ${}^{1}$School of Mathematics,
     Cardiff University,} \\
{\it Cardiff, United Kingdom} \\
{\it ${}^{2}$Department of Physics, Kinki University, } \\
{\it Higashi-Osaka, Japan} \\
}

\maketitle

\begin{abstract}
We review uniqueness theorems as well as other general results about higher dimensional
black hole spacetimes. This includes in particular theorems about the topology of higher dimensional spacetimes, theorems about their symmetries (rigidity theorem), and the classification of supersymmetric black holes. We outline the basic ideas underlying the proofs of these statements, and we also indicate ways to generalize some of these results to more general contexts, such as more complicated theories.
\end{abstract}

\section{Introduction}

The prediction of the existence of black holes by general relativity is one of the most intriguing ones in
all of theoretical physics, and there is now compelling evidence that such objects might indeed exist in Nature. Apart from thus describing, in all likelihood, real astrophysical objects, black holes are also viewed as a theoretical laboratory for ideas about the, not yet completely understood, laws of quantum gravity. These connections arise in particular through the analogy between the ``laws of black hole mechanics'', and the laws of thermodynamics~\cite{BardeenCarterHawking73}. They strongly suggest that black holes should have some sort of (quantum) statistical mechanical
description analogous to, say, the (quantum) statistical mechanical description of a gas comprised of atoms, see
e.g.~\cite{Wald01LR} for a review.

Both in the astrophysical context, as well as in the context of quantum gravity, the striking {\em uniqueness}, also referred to as {\em no hair property}, of stationary black hole solutions plays an important role: All (regular) stationary, asymptotically flat solutions of the Einstein-Maxwell equations in $D=4$ dimensions are uniquely determined by their mass, angular momentum, and electric charge, and are in fact given by the Kerr-Newman family of solutions. In the astrophysical context, this result is important because one expects that real physical black hole systems, while not being stationary as long as they form, will eventually settle down and thus eventually become stationary. Assuming this to be true, one can then restrict attention to the explicitly known Kerr-Newman family of solutions. Moreover, one can study small non-stationary perturbations of such a system within the well-understood framework of
linear perturbation theory, or by Post-Newtonian methods, possibly corrected by radiation-reaction effects, thus
eliminating, in a large dynamical range, the need to solve numerically the full set of non-linear Einstein-Maxwell
equations. In the quantum gravity context, one also relies in many approaches on the fact that the stationary black
hole solutions are functions of only a few conserved charges. While many classical properties of black holes are
probably lost in a full quantum regime, one still expects that such charges can be defined as quantum charge operators, that an ensemble of physical quantum states in the Hilbert space, characterized by a definite values of these parameters, can be identified, and that quantities such as entropy can be assigned to such an ensemble.

It is clearly of interest to ask whether a version of the black hole uniqueness theorem still holds under more general assumptions, for example (a) in the presence of other Einstein-matter fields systems such as Einstein-Yang-Mills, Einstein-scalar fields, Einstein-Euler, Einstein-Vlasov, Einstein-Skyrmions, etc. and (b) in more than four spacetime dimensions.

\medskip
\noindent
{\bf a) Inclusion of matter fields:}
As a rule of thumb, if the matter sector without coupling to gravity already contains soliton-like
static or stationary solution, then one has to expect that the corresponding theory coupled to Einstein-gravity
has hairy black holes. This is, in some sense, not really surprising, so it is generally more interesting to study
the uniqueness of systems in which the matter sector does not have soliton solutions in and of itself.
An interesting example of this type is Einstein-Yang-Mills theory. That theory does not have any soliton
like solutions without coupling to gravity, and neither has the vacuum Einstein theory. Nevertheless, for
suitable gauge groups ($\pi_2(G) \neq 0$), hairy black holes were found, to the surprise of many researches,
in the coupled theory. While the solutions that were originally found turned out to be unstable, there do exist
stable hairy black holes in other Einstein-matter theories, and furthermore, there are even hairy black holes which
do not have any kind of symmetry other than time-translations. Thus, for several Einstein-matter systems, the uniqueness theorems definitely fail to hold. For a detailed discussion of black hole uniqueness in $D=4$, including many references, see the review~\cite{Heusler96,CCH12}.

\medskip
\noindent
{\bf b) Higher dimensions:}
The possibility that there could exist more than the four macroscopically large, observed, dimensions
has been pursued almost from the beginning of general relativity theory. One of the main attractive features of this idea, common to most approaches, is that matter fields naturally ``arise from geometry''.
In string theory, extra dimensions appear naturally from constraints imposed by the quantum nature of strings propagating in a higher dimensional target space (usually $D=26$, or $D=10$). Particular values for the number of
dimensions also arise in supergravity theories due to the constraining nature of the supersymmetry algebra.
In practice, string theories are often analyzed within the supergravity approximation (especially in the context
of the AdS-CFT correspondence and its variants), hence by some sort of classical gravity theory in higher dimensional spacetime.

From a physical viewpoint one has to explain why, if extra dimensions indeed exist, we do not directly observe them. In the standard Kaluza-Klein approach, the extra dimensions are assumed to be extremely small (e.g. tori, or other compact manifolds with special properties, having small volume). Then, any fields with non-trivial dependence in the extra dimensions effectively have a very large mass (small deBroglie wavelength), and thus would decay into lighter fields in a quantum field theoretic description. There are, however, also other theoretical scenarios in which the extra dimensions are
not small, but macroscopically large. In this type of model, it is assumed that the four dimensions which we observe effectively constitute some timelike submanifold (``brane'') in a higher dimensional spacetime, to which the relevant matter fields (standard model) are confined in some way. In the classical context, one means by this usually that the contribution to the stress tensor by observable matter fields is localized on, or very near to, the brane. Apart from studying black hole solutions with boundary conditions at infinity corresponding
to the Kaluza-Klein respectively brane setup, one can also study higher dimensional black holes with standard asymptotically flat boundary conditions, i.e. which are asymptotic to $D$-dimensional Minkowski space (or e.g.
$D$-dimensional (Anti-)deSitter spacetime, where appropriate). The latter are motivated particularly in the context of holography, wherein a higher dimensional gravity theory with AdS-boundary describes a quantized field theory on the boundary. Asymptotically flat boundary conditions are not well-motivated physically in higher dimensions. However, they can serve as a clean setup to isolate new phenomena of higher dimensional gravity theories.

\medskip

In this review article, we will concentrate on uniqueness theorems in higher dimensional gravity theories, (b). Of course, because higher dimensional theories can always be reduced to lower dimensional ones with
additional matter fields via the Kaluza-Klein reduction process, (b) includes also many models in (a).
Not surprisingly therefore, much less is known about the uniqueness theory of higher dimensional black holes than
in 4 dimensions, in much the same way as much less is known about
general Einstein-matter systems compared to the vacuum theory in 4 dimensions.
Generally speaking, one might expect to be in
either of the following three, qualitatively different, situations:

\begin{enumerate}
\item (``Best possible case'') This would mean that all black hole solutions in the given theory (e.g. vacuum general relativity in $D$ dimensions, or including various special matter fields) would be specified uniquely by a finite set of ``conserved charges'' that are calculable as surface integrals over a large $(D-2)$-dimensional
    surface near infinity. For example, in pure Einstein-gravity, these would be the conserved ADM-type mass, and the ADM-angular momenta (or more restrictively even, as in $D=4$, the angular momenta corresponding to rotational Killing fields). In theories involving also matter fields such as Maxwell fields, these would include additionally e.g. electric/magnetic-type charges etc. It is clear that we cannot to be in this situation for general matter fields, because it is known even in $4$ dimensions that black hole uniqueness theorem in this form does not hold e.g. for the Einstein-Yang-Mills system, see (a). It is also clear that we cannot expect to be in this situation even for
    pure Einstein-gravity in $D$ dimensions, because solutions of $4$-dimensional Einstein-$SU(N)$-Yang-Mills theory are solutions of this theory via the Kaluza-Klein reduction process e.g. if the $D$-dimensional manifold is $\M \times SU(N)$, with $\M$ the 4-dimensional spacetime manifold. The latter solutions are not asymptotically flat from the point of view of the higher dimensional spacetime (they are asymptotically Kaluza-Klein, see below), so one might hope that black hole uniqueness in the sense of the ``best possible case'' might hold e.g. in pure Einstein-gravity in $D$ dimensions, if we restrict to stationary black holes that are asymptotically flat in the $D$-dimensional sense (i.e., asymptotic to $D$-dimensional Minkowski spacetime). However, this--to the surprise of many researchers--also turned out to be false when Emparan and Reall~\cite{ER02b} discovered 5-dimensional regular, stationary asymptotically flat black ``rings'' which had the same mass and angular momenta as the previously known Myers-Perry solutions~\cite{MP86}, but were not isometric to these and even had a different horizon topology (explicit forms of many higher dimensional black hole solutions are given e.g. in the review~\cite{ERLR}). Thus, we definitely cannot be be in the ``best possible case''.

\item (``Intermediate situation'') Since higher dimensional black holes cannot be uniquely specified by their conserved charges (ADM-mass, ADM-angular momenta, electric charge(s), $\dots$) alone, one might hope that they might be uniquely specified if one adds to these parameters further ones that somehow specify internal degrees of
    freedom of the stationary black holes that cannot be read off from the asymptotics at infinity via the conserved charges alone.
    A hint that this expectation might not be unreasonable comes from other systems of PDE's of geometrical nature
    admitting soliton-like solutions, but not containing the gravitational field. For example, it is well-known, that the space of self-dual solutions to the Yang-Mills equations $F = \star F$,
     modulo gauge transformations, on a sufficiently generic, compact,
    Riemannian $4$-manifold $\M$ is itself a manifold $X_k$ (see e.g.~\cite{selfdual}), whose dimension is given by $\dim \ X_k = 8k-3$ [for $SU(2)$], where $k=c_2(E)$ is the second Chern number of the
    associated bundle $E \to \M$, interpreted as the instanton number. The local coordinates on $X_k$
    have an interpretation in terms of the relative position, orientation, and size, of the instantons represented by the gauge field. Similarly, the space of $k$-centered BPS monopole solutions modulo gauge transformation on $\mr^3$ with gauge group $SU(N)$ is known to be a finite-dimensional manifold, see e.g.~\cite{AH}. The local coordinates of $X_k$ again have an interpretation in terms of the position, orientation, velocity and magnetic charges of the monopoles. Furthermore, in both examples, the dimension of the space of solutions modulo gauge transformations is related to the index of some differential operator.
    By analogy with these non-linear systems of PDE's, one may hope that the space of, say black hole solutions with
    $k$ components of the event horizon of given topology modulo gauge transformations, may be given by a manifold $X_k$ (or maybe more generally, a space with certain
    singularities) whose local coordinates describe the relative position, size, orientation etc. of the black hole horizons, and possibly other data encoding the topology/shape of the ambient manifold surrounding the black holes. One may further hope that the dimension of the space $X_k$ may be related to the index of some operator associated with the linearized
    Einstein equations. The present knowledge of black hole solutions in pure Einstein-gravity does not contradict this scenario, and it seems plausible that this should be the typical situation for supersymmetric solutions (in supergravities). However, there are no general results.

    From the classical viewpoint, proving scenario 2) would be no less satisfactory than 1). However, in the
    context of quantum gravity, the situation is less clear. While the conserved charges should retain their
    role also in quantum theory as operators whose eigenvalues are the charge quantum numbers of the corresponding quantum states, (as they do in ordinary quantum field theory), the role of the further ``internal parameters'' in scenario 2) at the quantum level is much less clear.

\item (``Worst case scenario'') The most general possibility is that the space of stationary black hole solutions, with, say, a fixed number of horizon components, is not finite dimensional in nature, i.e. that there are
    ``free function(s) worth'' of solutions with given asymptotic charges. In this case, there seems to be little point in trying to prove a ``uniqueness theorem''.
\end{enumerate}

In summary, while it is known that we are not in case 1), it is unknown whether we are in case 2) or 3), even
for asymptotically flat, stationary vacuum black holes in $D>4$ dimensions.

\medskip
\noindent
Related to the question of black hole uniqueness is that of stability. If a black hole is unstable, then one may expect that the endpoint of the evolution of an instability will be a new kind of black hole--assuming of course that the system settles down at all. Furthermore, stability analyses can in principle detect the existence of new stationary black holes which are perturbatively close to a given family of black holes. To understand the stability issue for a given family of stationary black holes, one may, for simplicity, start by looking at the linearized field equations around that background (although one should emphasize that linear stability does not necessarily imply stability against small {\em finite} perturbations). Unfortunately, already in the simplest cases such as a Schwarzschild black hole, the analysis of these equations is still rather involved~\cite{reggewheeler}. (See also \cite{KI03,IK03}. For a more modern treatment of the scalar version of the stability problem based on the method of ``energy norms'', see e.g.~\cite{dafermos}.) In the four dimensional Kerr case, there is available the Teukolsky formalism~\cite{teukolsky} which effectively reduces the full set of coupled perturbation equations to that of a single complex variable. Furthermore,
that equation can be studied via separation of variables techniques, and this analysis shows that the Kerr black hole is stable ``mode-by-mode''. Unfortunately, there is no known formalism of comparable power neither in higher dimensions, or in 4 dimensions with more general matter fields. Progress has been made (see~\cite{durkee}
and refs. therein) in higher dimensions for certain kinds of stationary black holes which are characterized by a wide separation of scales between various parameters characterizing the black hole, e.g. ``long-thin'' (highly prolate), or ``flat thin'' (highly oblate) horizons, partly based on numerical methods. Some of these black holes have thereby been determined to be unstable, and the nature of the instability is qualitatively similar to that known before
 for the 5-dimensional black string~\cite{gregory}. However, there is no complete picture. There are also other approaches to the black hole stability problem based on variational principles
related to ``local Penrose inequality'', or on the notion of thermodynamic stability, see~\cite{murata,HW}. Unfortunately, also these methods do not give a complete picture of stability of standard families of higher dimensional black holes.

Linearized analyses can also detect whether a new family of stationary black holes is ``branching off'' from a given family. Indeed, if a given black hole admits a stationary (non-gauge) perturbation, which is not tangent to the original family of solutions, then this is evidence--although not proof--for a new family of solutions. This has also been carried out in some examples~\cite{durkee}, and evidence has been found for the existence of a 70-parameter family of new stationary solutions in $D=9$ dimensions branching off from Myers-Perry black holes in a certain parameter range~\cite{Dias10}.
In particular, evidence has been found for families of solutions with a very small number of Killing vectors (i.e. with the minimum possible symmetry compatible with the rigidity theorem, see below). Evidence in the same direction has
also come from an entirely different approximation scheme, wherein the black object is treated as some sort of extended test-object, in much the same way as one can treat small objects as point-particles. This so-called ``blackfold approach''~\cite{blackfold} likewise indicates the existence of, yet unknown, asymptotically flat solutions with a very small number of Killing vectors. Thus, it appears from a variety of viewpoints that the manifold of black hole solutions in higher dimensions is largely unknown.

The plan of this article is as follows: We first outline, schematically, the proof of the black hole uniqueness theorem for Einstein-Maxwell theory in 4 dimensions. We then critically review to what extent the steps in this analysis generalize to higher dimensional theories, outlining the method of proof of the key intermediate theorems, which are also interesting in their own right as general structure theorems for higher dimensional black holes. We then present black hole uniqueness theorems in higher dimensions in the static, resp. stationary, case in higher dimensions. In the first case a rather complete picture emerges (in particular for supersymmetric solutions in certain supergravity theories), but in the second case, results are currently available only if
one imposes more symmetry on the solution than what one has the right to expect in view of the rigidity theorem and
also the numerical evidence. We will mostly restrict the discussion to either vacuum general relativity or simple extensions thereof, but we will try to indicate in what situation the techniques can be generalized to more general theories. Also, we will restrict attention mostly to asymptotically flat, or asymptotically Kaluza-Klein boundary conditions at infinity, and we will (except in the discussion of supersymmetric solutions) usually assume that the black hole is not extremal.

\section{General structure of uniqueness proof and history in $D=4$}

 For comparison with the higher dimensional case, we first give an outline of the proof that the Kerr-Newman black holes~\cite{Kerr63,Newman65} are the only stationary, asymptotically
flat black hole solutions in Einstein-Maxwell theory in $D=4$. The action is
\ben
S = \int_\M \frac{1}{2}R \ \star_4 1 + F \wedge \star_4 F \ .
\een
Here, $\M$ is a four dimensional spacetime manifold, with Lorentzian metric\footnote{Except in sec.~\ref{sec:susy},
our signature convention is $(-++\dots)$.} $g$ and the field strength is $F=\D A$.
It is assumed that $g,A$ satisfy the asymptotic conditions
\ben\label{afbndy}
g_{\mu\nu} = \eta_{\mu\nu} + O(r^{-1}) \ , \quad A_\mu = O(r^{-1})
\een
in a Cartesian coordinate system $x^\mu$ (and an appropriate gauge for $A$) that is defined in the asymptotic
region of $\M$. In particular, the asymptotic region
is required to be diffeomorphic to $\mr$ times $\mr^3 \setminus B^3_r$, with
$B^3_r$ a sufficiently large ball of radius $r$ with respect to the radial Cartesian coordinate $r=\sqrt{x_1^2+x_2^2+x_3^2}$.
As usual, it is also required that derivatives fall off by a correspondingly higher power.
The metric and gauge field are stationary in the sense that there exists a vector field $t$ on $\M$ which
is equal to $\partial/\partial x^0$ in the asymptotic region, such that
\ben
\pounds_t g = 0 = \pounds_t A \ .
\een
The asymptotic conditions allow one to construct the asymptotic null infinities $\I^\pm$ of the spacetime (see
the conformal diagram of Schwarzschild below\footnote{The extended Schwarzschild spacetime has two asymptotic ends, but in the rest of this paper we will restrict attention to one given asymptotic end of the spacetime under consideration.}), and the black hole may then be defined rigorously as the complement
of the past of future null infinity,
\ben
{\rm BH} = \M \setminus I^-(\I^+) \ , \quad
{\rm WH} = \M \setminus I^+(\I^-) \ ,
\een
and similarly the white hole. $I^\pm$ denote the chronological future/past of a set.
It is of course understood that BH region should not be empty.
As we will never be concerned with the interior of the black hole, we will normally use the symbol
$\M$ for the exterior of the
black and white hole, also called ``domain of outer communication'' (for a more precise
definition see sec.~\ref{sec1}). It corresponds to the blue region in the
figure. $\M$ has an inner boundary, $\partial
\M = \H$, comprised of the union of the future and past horizon $\H^\pm$. We wish to understand what
solutions $(A,g)$---subject possibly to further technical assumptions such as analyticity/global causal structure---there can be.

\begin{center}
\begin{tikzpicture}[scale=1.1, transform shape]
\shade[left color=blue] (0,0) -- (2,2) -- (4,0)  -- (2,-2) -- (0,0);
\draw (0,0) -- (-2,2) -- node[above left] {$\I^+$} (-4,0)  -- node[left] {$\I^-$} (-2,-2) -- (0,0);
\draw (4,0) node[right]{$i_0 \cong S^2$};
\draw (0,0) -- (2,2) -- node[above right] (c) {$\I^+$}  (4,0) -- node[right] (a) {$\I^-$} (2,-2) -- (0,0);
\shade[left color=gray] (0,0) -- (-2,2) decorate[decoration=snake] {-- (2,2)} --  (0,0);
\draw (0,0) -- (-2,2) decorate[decoration=snake] {-- (2,2)} -- (0,0);
\shade[left color=gray] (0,0) -- (-2,-2) decorate[decoration=snake] {-- (2,-2)} -- (0,0);
\draw (0,0) -- (-2,-2) decorate[decoration=snake] {-- (2,-2)} -- (0,0);
\draw[->, very thick] (2,3.2) node[above right]{singularity} -- (1, 2.2);
\draw[very thick, black] (0,0) -- (4,0);
\draw (2,0) node[below]{$\Sigma$};
\filldraw (0,0) circle (.05cm);
\filldraw (4,0) circle (.05cm);
\draw[->, very thick] (-2,3) node[above left] {BH $=\M \setminus J^{-}(\mathscr{I}^{+})$} -- (-.5,1.5);
\draw (1.3,1.3) node[black,above,left]{$\H^+$};
\draw (1,-1) node[black,below,left]{$\H^-$};
\draw(-0.2,0) node[black, left]{$B \cong S^2$};
\end{tikzpicture}
\end{center}

The analysis proceeds in the following steps (we restrict ourselves
for simplicity to non-extremal black holes, $\kappa > 0$, see eq.~\eqref{surfgr}):

\begin{enumerate}
\item Using the geometric nature of the horizon $\H$, and the field equations, one shows~\cite{H72,HE} that each horizon cross section is topologically a sphere, $B \cong S^2$. This is called the ``topology theorem''. In particular $\H^\pm \cong \mr \times S^2$.
\item Now consider the restriction of $t$ to the horizon, $\H=\H^+ \cup \H^-$. Then we can be in either of the following
cases: (a) The horizon is non-rotating, in the sense that $t$ is null on $\H$, i.e.
tangent to the null-generators. Then the ``staticity theorem''
of~\cite{SW92} implies that the spacetime is not only {\em stationary} but even {\em static}. This means
that there is a foliation by Cauchy surfaces $\Sigma(\tau)$ of $\M$ intersecting the bifurcation surface $B$ such that $t$ is orthogonal to each $\Sigma(\tau)$. In fact, we may chose $\tau$ as one of the coordinates, and then,
identifying $\Sigma(\tau)$ with $\Sigma$ along the orbits of $t$, we can write $\M = \mr \times \Sigma$,
\ben
g = -N^2 \ \D \tau^2 + h \ , \quad A = \phi \ \D \tau \ , \quad t = \frac{\partial}{\partial \tau}
\label{metric:static}
\een
where $h$ is a Riemannian metric on $\Sigma$, $N>0$ a smooth function (lapse), and $\phi$ is the electro static
potential, all independent of $\tau$.

(b) On the other hand, if $t$ is not tangent to the generators on $\H$, then the horizon is rotating.
Using the topology theorem $B \cong S^2$, and that a vector field on $S^2$ necessarily vanishes somewhere, one can then establish the ``rigidity theorem''~\cite{HE}, under the assumption of analyticity of the
spacetime. This theorem states that there exists a second Killing field $\psi$
on $(\M, g, A)$ (i.e., both $g$ and $A$ are Lie-derived by $\psi$) commuting with $t$ such that the flow generated by $\psi$ has closed orbits with period, say, $2\pi$, and such that
\ben
K = t + \Omega \ \psi \ ,
\een
is tangent to the null-generators of the horizon. In particular, the horizon is a ``Killing horizon,'' whose surface
gravity $\kappa > 0$, defined through
\ben\label{surfgr}
\nabla_K K = \kappa \ K
\een
can then be shown\footnote{The proof of this statement only requires that the Einstein
equation holds with a stress tensor satisfying the dominant energy condition.} to be constant on $\H$~\cite{Wald84}.
The constant $\Omega$ is interpreted ``angular velocity'' of the horizon.
\end{enumerate}

At this stage, the argument branches off, depending on whether one is in the non-rotating (i.e. static) case (a), or the rotating case (b).

\begin{enumerate}
\item[3a)]
\underline{Static case :} In the static case, there are now several arguments that the solution under consideration must be the (regular, non-degenerate) Reissner-Nordstrom solution, characterized uniquely by its mass $M$ and electric charge $Q$. Traditionally, one proceeds via Israel's theorem~\cite{Israel67wq} which states that the solution must
be rotationally symmetric, i.e. invariant under $O(3)$. The rest is a straightforward integration of Einstein's
equation subject to staticity and $O(3)$-symmetry, as is done in Schwarzschild. A more modern version is
given in~\cite{Ruback88}\footnote{Note that his argument contained a gap.} and~\cite{BM,Masoodulalam2}.
The method of the last paper is particularly elegant and consists of
the following steps, where we restrict for simplicity the discussion to the vacuum case, $A=0$. One starts by
doubling the spatial slice $(\Sigma,h)$ across the horizon cross section, gluing a copy $\overline \Sigma$ onto $\Sigma$
along $B$. One then performs a conformal rescaling from $h$ to $\tilde h = \Omega^2 h$
on the doubled spacetime $\tilde \Sigma = \Sigma \cup \overline \Sigma$, with $\Omega$ chosen in such a way that (a), the scalar curvature $\tilde R$ of $\tilde h$ is non-negative, (b) such that $\Omega \to 1$ near the
spatial infinity $i_0$ of $\Sigma$, and (c) such that $\Omega \to 0$ near the spatial infinity $\overline i_0$
of $\overline \Sigma$, in such a way that $\overline \Sigma$ is effectively compactified, or ``capped off''. (d)
The mass $\tilde M$ of $(\tilde \Sigma, \tilde h)$ is zero.
Then the positive mass theorem is applied to $(\tilde \Sigma, \tilde h)$, implying that
$\tilde h$ is in fact the flat Euclidean metric on $\tilde \Sigma = \mr^3$. The rest of the proof then proceeds by showing that $\Omega$
must be precisely so that $\Omega^{-2} \delta= h$ is equal to the spatial part of the Schwarzschild metric
in isotropic coordinates, from which it then easily follows that the spacetime itself is Schwarzschild. We will give a more detailed account of this argument, taken from~\cite{Gibbons02b} and valid also for higher dimensions and various matter fields, in sec.~\ref{sec:static}.

\item[3b)]
\underline{Rotating case:} Here one proceeds as follows (again we restrict attention for simplicity to $A=0$).
First, using Einstein's equation, one shows that the metric on $\M$ can be written {\em globally} in Weyl-Papapetrou
form, meaning that {\em globally}, coordinates $r>0,z,\tau$, together with a $2\pi$-periodic coordinate $\varphi$
can be chosen, in such a way that
\ben
g =   - \frac{r^2 \, \D \tau^2}{f} + \e^{-\nu} (\D r^2 + \D z^2)
+
f( \D \varphi + w \, \D \tau)(\D \varphi + w \, \D \tau)\ , \quad \psi= \frac{\partial}{\partial \varphi} \ ,
\quad t = \frac{\partial}{\partial \tau} \ .
\een
The remaining components of the metric, $f,\nu, w$ are functions of $r>0,z$ only, and they obey equations that follow from the original Einstein equations. Assume now that there were two solutions with equal values of the mass $M$ and angular momentum $J$ relative to the Killing field, defined as
\ben
J = \frac{1}{8\pi} \int_\infty \star_4 \ \D\psi
\een
where the integral is over a large sphere in the asymptotic region. Then, one derives, using the Weyl-Papapetrou
form that both metrics must satisfy, a partial differential equation in $r,z$ for a quantity measuring the
``difference'' between non-trivial metric components of the two metrics. The particular form of this identity,
called ``Robinson identity''~\cite{robinson} allows one to prove that this ``difference'' actually has to be zero, and hence that the metrics actually have to coincide. It was originally not clear how to generalize the Robinson identity to
the case of Einstein-Maxwell theory in 4 dimensions, but this problem was later solved by~\cite{Mazur84} and
\cite{Bunting83}.
Their work in particular showed that the Robinson identity can be understood from the point of view of certain
non-linear sigma-models. It is this viewpoint that proves useful also for other theories, and in higher dimensions, and we will review it in some more detail below in sec.~\ref{sec:sigma}.
\end{enumerate}

\section{Higher dimensions}

In this section, we will critically investigate the various steps described above in 4 dimensional Einstein Maxwell
theory, and see to what extent they can be generalized to higher dimensions. We will then explain what kinds
of uniqueness theorems are presently available. For simplicity, we will mostly
restrict to vacuum general relativity, and to non-extremal black holes
[i.e.  non-vanishing surface gravity, see eq.~\eqref{average:surfacegravity}],
but we will also comment on more general cases, mostly in sec.~\ref{sec:other}.

\subsection{Asymptotic conditions}

\label{sec1}
Let $(\M,g)$ be a $D$-dimensional, stationary black hole spacetime $D \ge 4$. The asymptotically timelike Killing field is called $t=t^\mu \partial_\mu$, so $\pounds_t g = 0$. Depending on the theory under consideration, or the nature of the solutions
that one is interested, one may wish to impose different asymptotic conditions on the metric and corresponding
conditions on the
matter fields, if those are present in the theory. Standard asymptotic conditions on the metric are:

\begin{enumerate}
\item Asymptotically flat boundary conditions: The metric approaches $D$-dimensional
Minkowski spacetime $\mr^{D-1,1}$ at large distances as in eq.~\eqref{afbndy}, with $O(r^{-1})$ replaced by
$O(r^{-(D-3)})$,  or at late advanced/retarded time (``null infinity'').
Note that the asymptotic symmetry group of an asymptotically flat spacetime is $SO(D-1,1) \times \mr^D$
(semi-direct product). In particular, the number $N$ of commuting Killing fields with circular orbits
must correspond to a subgroup $U(1)^N \subset SO(D-1,1)$ of the Cartan subalgebra. Hence,
\ben
N \le \left\lceil \frac{D-1}{2} \right\rceil \ .
\een
Here $\lceil x \rceil$ is by definition the largest integer $n \le x$.

\item If the theory has a negative cosmological constant, or scalar fields with a minimum of the potential that effectively provides this, then asymptotically Anti-deSitter (AdS) boundary conditions are appropriate, see e.g.~\cite{HIM}, or many other references. These boundary conditions can be viewed as saying that a suitably conformally rescaled spacetime $\Omega^2 g$ has
    a timelike conformal boundary. Similarly, if the theory has a positive cosmological constant, then
    it is reasonable to consider asymptotically deSitter ``boundary'' conditions. In that case, one has a
    spacelike conformal boundary. We will not discuss here either type of theory, although they are of considerable interest and various exact black hole solutions have been found. However, especially in the AdS case, recent
    investigations indicate that the manifold of black holes may be very complicated~\cite{Dias,bizon}.

\item Asymptotically Kaluza-Klein (KK) boundary conditions (see below): The spacetime is asymptotically
$\mr^{s,1} \times Y^{D-s-1}$, where $Y^{D-s-1}$ is a compact Riemannian manifold ($D-s-1$ extra-dimensions). Alternatively,
one can consider the asymptotics $AdS_{s+1} \times Y^{D-s-1}$ or $dS_{s+1} \times Y^{D-s-1}$.
Again, if we have $N$ commuting circular Killing fields, then $N$ is restricted by the number of
circular symmetries that $\mr^{s,1} \times Y^{D-s-1}$ has, and similarly in the $dS$ and $AdS$-cases. The maximum $N=D-3$ can be achieved if $s=1,2,3,4$, and $Y^{s-D-1} \cong \T^{D-s-1}$. We will exclusively deal with this case in this review.

\item Other variants of the above asymptotic conditions can also be considered. For example, one might
replace the asymptotically flat condition by ``locally asymptotically flat''. This means that
the spacetime is not asymptotically $\mr^{D-1,1}$, but instead $\mr^{D-1,1}/\Gamma$, where $\Gamma \subset O(D-1)$ is some discrete subgroup of the spatial rotations. For example, in $D=5$, we could take a cyclic subgroup $\Gamma=\mz_p \subset O(4)$. Then the large
spheres $S^3$ near spatial infinity are replaced by a quotient $S^3/\mz_p$. Of course, one can consider also other discrete subgroups, KK-quotients, etc. Black hole spacetimes of this nature have been given e.g. by~\cite{teo}.

\item One may also study ``braneworld boundary conditions''; such black holes have recently been found
numerically~\cite{FLW11,FW11}. We will not consider them here.
\end{enumerate}
Asymptotically KK-boundary conditions are in more detail as follows:
We assume that a subset  of $\M$ is diffeomorphic to the cartesian product of
$\mr^s$ with a ball removed---corresponding to the asymptotic region of the large spatial
dimensions---and $\mr \times \T^{D-s-1}$---corresponding to the time-direction and small dimensions.
We will refer to this region as the asymptotic region and call it $\M_\infty$.
The metric is required to behave in this region like
\ben\label{standard}
g = -\D\tau^2 + \sum_{i=1}^s \D x_i^2 + \sum_{i=1}^{D-s-1} \D \varphi_i^2 + O(R^{-s+2}) \, ,
\een
where $O(R^{-\alpha})$ stands for metric components that drop off at least as fast as $R^{-\alpha}$ in the radial
coordinate $R = \sqrt{x_1^2+...+x_s^2}$, with $k$-th derivatives in the coordinates
$x_1, \dots, x_s$ dropping off at least as fast as
$R^{-\alpha-k}$. These terms are also required to be independent of the coordinate
$\tau$, which together with $x_i$ forms the standard cartesian coordinates on $\mr^{s,1}$.
The remaining coordinates $\varphi_i$ are $2\pi$-periodic and parameterize the torus
$\T^{D-s-1}$. The timelike Killing field is assumed to be equal to $\partial/\partial \tau$ in
$\M_\infty$. We call spacetimes satisfying these properties
asymptotically Kaluza-Klein spacetimes\footnote{For the axisymmetric spacetimes considered in
this paper, we will derive below  a stronger asymptotic expansion, see~\cite{HollandsYazadjiev08b}}.

The domain of outer communication is defined by
\ben
\langle \! \langle \M \rangle \! \rangle =
I^+ \left(
\M_\infty
\right) \cap
I^- \left(
\M_\infty
\right)\,,
\een
where $I^\pm$ denote the chronological past/future of a set. The black
hole region $B$ is defined as the complement in $\M$ of the causal past of the
asymptotic region, and its boundary $\partial B = \H$ is called the (future)
event horizon. Since we will never be concerned with the interior of the black hole,
we will simply write $\M$ again for the domain of outer communication.

In this paper, we also sometimes assume the existence of $D-3$ further linearly independent Killing fields,
$\psi_1, \dots, \psi_{D-3}$, so that the total number of Killing fields is
equal to the number of spacetime dimensions minus two. These are required to mutually
commute, to commute with $t$, and to have periodic orbits.
The Killing fields $\psi_i$ are referred to
as ``axial'' by analogy to the four-dimensional case, even though their
zero-sets are generically higher dimensional surfaces rather than ``axis''
in $D>4$, see the discussion below in sec.~\ref{ref:wp}. We also assume that, in the asymptotic region $\M_\infty$, the action of the axial symmetries
 is given by the standard rotations in the cartesian product of
flat Minkowski spacetime $\mr^{s,1}$ times the standard flat torus $\T^{D-s-1}$.
In other words, $\psi_i = \partial/\partial \varphi_i$ or\footnote{The
notation $\lceil x \rceil$ means the largest integer $n$ such that $n \le x$.} $\psi_j = x_{2j-1} \partial_{x_{2j}}
- x_{2j} \partial_{x_{2j-1}}$ for $j=1, \dots, \lceil s/2 \rceil$
in $M_\infty$. The group
of isometries is hence $G = \mr \times K$, where $\mr$ corresponds to the flow of
$\tau$, and where $K = \T^{D-3}$ corresponds to the commuting flows of the axial
Killing fields.

\medskip

Unfortunately, in order to make many of the
arguments in the following sections in a consistent way, one has to make certain further technical
assumptions about the global nature of $(\M,g)$ and the action of the symmetries.
Our assumptions are in parallel to those made by Chru\'sciel and Costa
in their study~\cite{CC08} of 4-dimensional stationary black holes. The
requirements are (a) that $\M$ contains an acausal, spacelike, connected
hypersurface $\Sigma$ asymptotic to the $\tau = 0$ surface in the asymptotic region $\M_\infty$,
whose closure has as its boundary $\partial \Sigma = B$ a cross section
of the horizon. We always assume $B$ to be compact but we allow for multiple components.
(b) We assume that the orbits of $t$ are complete. (c) We assume that
the horizon is non-degenerate. (d) We assume that $\M$
is globally hyperbolic. We will also occasionally assume (e) that the spacetime, the metric, and
the group action are analytic, rather than only smooth.

\subsection{Rigidity theorem}\label{sec:rigidity}

The rigidity theorem is not only a key ingredient in the black hole uniqueness theorems, but also
important on its own right. This is because it shows that every stationary black hole horizon is, in fact, a Killing horizon,
i.e. that there is a Killing field $K$ tangent to the generators of $\H$ commuting with $t$, or coinciding with it.
That in turn implies~\cite{Wald84} the constancy of the surface gravity, which otherwise would not even be defined.
Because the constancy of the surface gravity is physically interpreted as the constancy of the temperature of the black hole (zeroth law of black hole mechanics), this result is of fundamental physical importance, and it is also the basis for the other laws of black hole mechanics. The precise statement of the rigidity theorem~\cite{HIW07,MI08} is as follows:

\begin{thm}\label{thm:3.1}
(``Rigidity theorem'') Let $(\M,g)$ be an asymptotically flat, analytic stationary
black hole solution to the vacuum Einstein equations.
Assume further that the event horizon, $\H$, of the black hole is
analytic and is topologically $\mr \times B$,
with $B$ compact and connected, and that the average surface gravity
$\langle \kappa \rangle \neq 0$ (see eq.~(\ref{average:surfacegravity}) below).
Then there exists a Killing field $K$, defined in a region that
covers $\H$ and the entire domain of outer communication, such that
$K$ is normal to the horizon and $K$ commutes with $t$.
\end{thm}

\noindent
The rigidity theorem has the following consequence~\cite{HIW07,MI08}:

\begin{thm}\label{thm:3.2}
Under the same assumptions made in Theorem~3.1 above,
if $t$ is not tangent to the generators of $\H$, then there exist
mutually commuting Killing fields
$\psi_1, \dots, \psi_N$ (with $\lceil \frac{D-1}{2} \rceil \ge N \ge 1$) with closed orbits
with period $2 \pi$ which are defined in a region that covers $\H$
and the entire domain of outer communication.
Each of these Killing fields commutes with $t$, and
\ben
  t = K + \sum_{i=1}^N \Omega_i \ \psi_i
\een
for some constants $\Omega_i \neq 0$, all of whose ratios are irrational.
\end{thm}

\noindent
The proof of both results relies heavily on the Einstein equations and also on some global results such as
topological censorship, see below.
First, one shows that near $\H$, the spacetime
$\M$ is doubly foliated by a 2-parameter family $B(u,r)$ of compact cross sections, such that the metric takes the
``Gaussian null form''~\cite{Penrose,MI83}
\ben\label{gaussian}
g = 2 \ \D u( \D r - r\alpha \ \D u - r\beta_a \ \D x^a) + \gamma_{ab} \ \D x^a \D x^b \ .
\een
Here, $x^a$ are local coordinates on $B$, and $\alpha, \beta = \beta_a \D x^a, \gamma = \gamma_{ab} \D x^a \D x^b$ are a scalar field, 1-form, and Riemannian metric on each of the spheres $B(u,r)$ that are parameterized by $u,r$. The horizon $\H$ is at $r=0$.
The form of the metric implies that $\ell = \partial/\partial r$ and $n = \partial/\partial u$ are commuting
vector fields. $\ell$ is a null vector field transverse to $\H$ which is also tangent to a congruence of null geodesics. $n$ is null on $\H$, and on $\H$, tangent to a congruence of null-geodesics, therefore
\ben\label{nnaln}
\nabla_n n = \alpha \ n \qquad \text{on $\H$.}
\een
``Gaussian null coordinates'' are adapted particularly well to the geometry near a null surface.
The Ricci tensor of $g$ can be expressed in terms of the fields $\alpha, \beta, \gamma$; one obtains (see e.g.~\cite{HIW07}):
\bena
R_{ab} &=&- \pounds_\ell\pounds_n \gamma_{ab}
         - \alpha \pounds_\ell \gamma_{ab}
         + R_{ab}(\gamma)
         - \delta^c{}_{(a} \delta^d{}_{b)} D_c\beta_d
         - \half \beta_a\beta_b+O(r)\non\\
R_{ua} &=&- D_a \alpha
  + \half\pounds_n\beta_a
  + \quater \beta_a \gamma^{bc} \pounds_n \gamma_{bc}
  - \delta^d{}_{[a}\delta^e{}_{b]}D_d(\gamma^{bc}\pounds_n \gamma_{ce})+O(r)\\
R_{ur} &=&-2 \pounds_\ell\alpha
         + \quater \gamma^{ca}\gamma^{db}(\pounds_n\gamma_{cd})\pounds_\ell\gamma_{ab}
         -\half \gamma^{ab} \pounds_\ell \pounds_n \gamma_{ab}
         -\frac{1}{2} \alpha\: \gamma^{ab} \pounds_\ell\gamma_{ab}
         -\half \beta_a\beta^a + O(r)\non
\eena\label{gee}
Here we have not written out for simplicity terms of order $O(r)$ that vanish on $\H$, but the form of
these terms is needed in the proof of the rigidity theorem (see e.g.~\cite{HIW07}). The $uu,ar$ and $rr$ components
are also not written, but are likewise needed.
If $\H$ was already known to be a Killing horizon, then we could choose the foliation $B(u,r)$ in such a way
that $n = K$, and hence, in view of eq.~\eqref{surfgr}, that $\alpha = \kappa = cst.$ Even though we do not know this at this stage, the
{\em average surface gravity} may be defined by
\ben
\label{average:surfacegravity}
\langle \kappa \rangle = \frac{1}{{\rm vol}(B)} \int_B \alpha \ \D S \ ,
\een
where $\D S$ is the volume element associated with $\gamma$, and where $B$ is any
of the surfaces $B(u,r=0)$. The average surface gravity turns out to be independent of $u$, because
$\alpha$ and in fact any of the tensor fields $r,u,\alpha,\beta,\gamma$ is Lie-derived by $t$.

The desired Killing vector field, $K$, is constructed in two steps.
\begin{enumerate}
\item First it is constructed locally in a neighborhood of $H$, in the sense that
the `Taylor expansion' around $\H$ satisfies
\ben
 \underbrace{\pounds_\ell \, \pounds_\ell \, \cdots \,
 \pounds_\ell}_{m \,\, {\rm times}}
(\pounds_K \Phi) = 0 \,, \quad m=0,1,2, \dots
\quad \mbox{on $\H$} \,,
\label{coeff:Taylor}
\een
where $\Phi$ denotes either one of the fields $\alpha, \beta, \gamma$.
\item Then $K$ is extended to the domain of outer communication.
For the purpose of Step 2), we assume that the spacetime metric and
matter fields be real analytic. The fact that $\M$ is simply connected (see sec.~\ref{sec:top})
implies that the extensions are single valued tensor fields on $\M$.
\end{enumerate}
Thus, if we assume analyticity, the main work of the proof is step 1).
The key idea of the proof is to choose $K := n$ {\em for a suitable choice} of the foliation $B(u,r)$--to be determined!--such that the metric takes Gaussian null form. Note that, since $u,r$ are affine parameters by construction, the foliation is determined uniquely once we give an arbitrary member $B = B(u,r=0)$, and on $B$, we choose
$n, l$ subject to $1=g(n,l)$.

To start, one shows that automatically $\pounds_n \gamma = 0$, for {\em any
foliation} of the type described. This is proved via the Raychaudhuri equation
\ben
\frac{\D}{\D \lambda} \theta_n = -\frac{1}{D-2}\theta_n^2 - \sigma_n^2 - 8\pi \ T_{uu}^{}
\een
where we mean the expansion $\theta_n$ and shear $\sigma_n$ of the congruence generated by $n$.
The stress energy term $T_{uu}$ actually vanishes due to the vacuum Einstein equations. With the help of this
equation, and the global structure of $\H$, one now derives, as in the area theorem, that $\theta_n = 0 =
\sigma_n$, and because these tensor fields are on $\H$ the trace and trace-free parts of $\pounds_n \gamma$,
we are done. However, from the $ua$ component of Einstein's equations written out in Gaussian null-coordinates~\eqref{gee} we then get:
\ben
\half \pounds_n \beta = \D \alpha \ , \qquad \text{on $B$},
\een
and because $\alpha$ is not constant on $B$ unless we choose a very special foliation,
we see that $\pounds_n \beta \neq 0$. Thus, for a general foliation,
condition~\eqref{coeff:Taylor} already fails for $m=0$, in the case when $\Phi = \beta$. We wish to find a
special foliation where this does not happen.

For a given foliation, we may decompose
$t = n + s$, where $s$ is tangent to each $B(u,r)$. A key point is that, while the Killing vector field
$t$ is given,  $n \to \tilde n$ and $s \to \tilde s$ individually change if we change the foliation to
$\tilde B(\tilde u, \tilde r)$.
It turns out that to find the desired foliation $\tilde B(\tilde u, \tilde r)$,
with corresponding tensor fields $\tilde u, \tilde r, \tilde \alpha, \tilde \beta, \tilde \gamma$,
one has to integrate along the orbit of $s$
two ordinary differential equations on $B$, for the quantities $f = \D \tilde u/ \D u$
respectively $\tilde u$ on $B$. These differential equations are
\ben\label{diffequns}
\pounds_s \log f = \alpha \ , \quad \pounds_s \tilde u = 1-\frac{1}{f} \ .
\een
It furthermore turns out that, once $\tilde u$, hence $\tilde r, \tilde \alpha, \tilde \beta, \tilde \gamma$, have been determined through~\eqref{diffequns}, the remaining conditions~\eqref{coeff:Taylor} are automatically satisfied for {\em all} $m$. The non-obvious proof of this uses Einstein's equations in the form~\eqref{gee}, but expanded to {\em all} orders in $r$.
Thus, what remains to be solved is eqs.~\eqref{diffequns}, both of which are schematically of the form
\ben
  \pounds_s \Psi = J \ .
\label{eq:*}
\een
When solving this equation with respect to $\Psi$,
the spacetime dimension and the topology of $B$ plays a crucial role.

For $4$-dimensions, the cross-section $B$ must be topologically a
$2$-sphere due to Hawking's topology theorem~\cite{H72}. It then immediately follows
that the flow of $s$ on $B$ must have a fixed point, $p$,
as the Euler characteristic of $B \cong S^2$ is non-zero.

Now the (infinitesimal) action of $s$ on any $2$-vector,
$v$, on the tangent space at the fixed point $p$ (where
$s=0$), is $(\pounds_s v)^a = - v^b(D_b s^a)$.
Since $D_b s^a$ is a linear map on the tangent space $T_p B$, which is
anti-symmetric with respect to $\gamma_{ab}$,
the action of $s$ describes an infinitesimal `rotation'
on the tangent space. Therefore all the orbits of $s$
must be closed with a certain period $P$.
Then, by integrating Eq.~(\ref{eq:*}) along a closed orbit of $s$ one can
always find a well-defined solution $\Psi$ which determines uniquely
our desired foliation $\tilde B(\tilde u, \tilde r)$. Putting $K = \tilde n$,
one can furthermore see inductively that eqs.~\eqref{coeff:Taylor} are satisfied for any
number of Lie derivatives by taking multiple $\pounds_l$ derivatives
of~\eqref{gee}\footnote{Here one also has to use the explicit form
of the $O(r)$ terms, as well as the remaining Einstein equations.}. Thus, one has completed step 1). Step 2) is accomplished, as already mentioned, by analytic continuation.

\medskip

In higher dimensions $D>4$, however, cross-sections $B$ of
the event horizon can admit non-trivial topology, and there is no reason
that the isometries of $B$ generated by $s$ need to have closed
orbits even if $s$ vanishes at some point $p \in \Sigma$.
(This would be the case even in $4$-dimensions if the horizon
topology could be e.g. a torus).
An example is supplied by considering a $5$-dimensional
Myers-Perry black hole solution~\cite{MP86},
whose event horizon cross-section is topologically $B \cong S^3$.
The solution admits two rotational Killing fields,
$\psi_1, \psi_2$ and their linear combination
provides $s=\Omega_1 \psi_1 + \Omega_2 \psi_2$ on $B$.
If we choose two rotational parameters in the linear combination
so that their ratio becomes incommensurable--i.e. $\Omega_1/\Omega_2 \notin {\mathbb Q}$--then the orbits of
$s$ are not closed on $B$. In this case, the argument which works in $D=4$ simply does not work as stated in higher dimensions.

Let us see in more detail what is the potential issue with eq.~(\ref{eq:*}),
and how we can overcome it. First, it turns out that also in higher dimensions,
we can also in general always decompose $s$ as
$s = \sum_i \Omega_i \psi_i$ for $N$ commuting Killing vector fields
$\psi_i$ on $(B, \gamma)$ with closed orbits, and corresponding flow
(an action of the torus ${\mathbb T}^N$ on $B$)
$F_{\underline \tau}$, with $\underline{\tau} = (\tau_1, \dots, \tau_N)$
a vector of parameters corresponding to the commuting Killing fields.
For each fixed $x \in B$, consider the Fourier transform (with $\underline m \in {\mathbb Z}^N$)
in the variables $\tau_i$:
\ben
\widehat{J} (x, \underline{m}) = \frac{1}{(2\pi)^N}
\int_{\mr^N} \ J[F_{\underline \tau}(x)] \ \e^{i\underline \tau \cdot \underline{m}} \ \D^N \tau  \ .
\een
Then a {\sl formal} solution $\Psi$ to \eqref{eq:*} is given by
\ben
\Psi(x)
  = i\!\sum_{m_1, ..., m_N \in {\mathbb Z} \setminus \underline 0} \!
    \frac{\widehat{J}(x,\underline{m})}{\underline{m} \cdot \underline{\Omega}}
  = i\!\sum_{m_1, m_2 \in \mz \setminus \underline{0}} \!
     \frac{ \widehat{J}(x,m_1,m_2)
          }{m_1\Omega_2}
     \cdot \Bigg( \frac{\Omega_1}{\Omega_2} - \frac{m_2}{m_1} \Bigg)^{-1} \,.
\label{sol:2torus:Psi}
\een
In the second line, we have restricted to the case $N=2$, to illustrate the potential problem with this expansion.
The point is that any irrational number, $\Omega_1/\Omega_2 \notin {\Bbb Q}$,
can be approximated by rational numbers $ m_2/m_1
\in {\Bbb Q}$ arbitrarily closely, by taking
$m_1,\: m_2, \rightarrow \infty$ in a suitable manner.
This implies that the denominator of the right side of the above equation can
become arbitrarily small and therefore that the series for $\Psi$ might not be convergent.

Nevertheless, this difficulty has been overcome
for non-extremal black holes (i.e., the case in which $ \langle \kappa \rangle \neq 0$) by employing a novel approach, and the rigidity result has thereby been
generalized to higher dimensions by~Refs.~\cite{HIW07,MI08}.
A key new idea employed in Ref.~\cite{HIW07} is to apply basic
results from {\em ergodic theory} (see e.g.~\cite{Walter}) to the flow $F_{\underline{\tau}}$
under consideration. To be able to apply these results, it is important here that $B$ is compact, and that the flow is
 isometric, i.e. in particular area preserving. When combined with Einstein's equations,
 the results from ergodic theory enable one to solve the desired differential equations~\eqref{diffequns}
 without appealing to the explicit series solution above, and hence avoiding the potential `small denominator problem'.

Unfortunately, for the extremal black hole case $\langle \kappa \rangle =0$, these methods
using ergodic theory do not seem to generalize in a straightforward manner.
But if we additionally require  (in the example of $N=2$-dimensional torus above),
that there exists a $q>0$ such that
\ben
\label{condi:dioph}
 \Bigg| \frac{\Omega_1}{\Omega_2} - \frac{m_2}{m_1}  \Bigg|
 > \frac{1}{m_1^q}
\ ,
\een
then one can show that the formal series solution, $\Psi$, to
\eqref{eq:*}, given by eq.~(\ref{sol:2torus:Psi}), is convergent
and therefore well-defined, and that it gives the desired (analytic) solution
to our equation.
Condition~\eqref{condi:dioph}---called {\em Diophantine condition}---is known to be satisfied for some $q$ except when
$\Omega_1/\Omega_2$ happens to be in a special set of irrational numbers of  measure zero. Therefore,
we can virtually always solve eq.~(\ref{eq:*})~\cite{HI09}. A similar
condition can also be formulated for $D \ge 6$, see~\cite{HI09}. Again, this generalized
Diophantine condition holds for all $\underline \Omega \in \mr^N$ except for a set of Lebesgue
measure zero.

\medskip
With the additional condition of the type~\eqref{condi:dioph},
the rigidity theorems above have been extended to include
extremal black holes in Theorems~1 and~2 of Ref.~\cite{HI09}.
Note that when $N=1$,
the Diophantine condition is automatically satisfied. But
when $N>1$---which can happen only in higher dimensions---the condition
is non-trivial. In this sense, the theorems for the extremal black hole
case are weaker than the theorems for the non-extremal case.

\medskip
A few remarks on the rigidity theorems for both extremal and non-extremal
cases are in order:
\begin{enumerate}
\item
Theorems~\ref{thm:3.1} and~\ref{thm:3.2} above, and those corresponding to the extremal case in
Ref.~\cite{HI09} hold also true for stationary black holes
coupled to matter fields in a fairly general class of theories
that include scalar fields taking values in some Riemannian target space $(X, {\mathcal G})$,
Abelian gauge fields with or without a Chern-Simons type term in the action, as well as cosmological constant, described by an
action of the form
\ben\label{genth}
S = \int_\M \frac{1}{2} R \ \star_D 1 + \frac{1}{2} {\mathcal G}_{ab}(\phi) \
\D \phi^a \wedge \ \star_D \D \phi^b + {\mathcal V}(\phi) \ \star_D 1 +
{\mathcal H}_{AB}(\phi) F^A \wedge \ \star_D F^B \ ,
\een
plus possibly a Chern-Simons term.
The key requirements on ${\mathcal H}, {\mathcal V}$ are essentially that the stress
tensor component in the Raychaudhuri equation
satisfies the energy condition $T_{uu} \ge 0$, which means that ${\mathcal H}_{AB}$ and ${\mathcal G}_{ab}$
should be positive definite. Thus, the above theorems in particular apply to stationary,
asymptotically (anti-)de Sitter black holes as well.
\item
The theorems apply not only to a black hole horizon but also to any horizon
defined as a ``boundary'' of the causal past of a complete
orbit of some Killing vector field, such as
a cosmological horizon if it exists.
\item
One can partially remove the analyticity assumption for the black hole
interior, following the strategy of Refs.~\cite{FRW99,Racz00}.
For the non-extremal case, the event horizon is isometric to a portion
of some bifurcate Killing horizon~\cite{RW92,RW96}. Then one can use
the bifurcate horizon as an initial data surface for $K$ defined in
a neighborhood of $\H$. Then, applying a characteristic initial value
formulation to extend $K$ into the interior of the black hole.
This type of characteristic initial value problem is ill-defined
towards the black hole exterior region and therefore would not appear
to be applicable to extend $K$ into the domain of outer communication.
Nevertheless,  remarkable progress has recently been made along this
direction~\cite{IK09a,IK09b} in $D=4$. Unfortunately, these methods
depend at the moment on the consideration of a tensor field (``Mars Simon tensor''~\cite{mars})
which has remarkable properties in $D=4$, but which has no obvious generalization
to higher dimensions. Thus, it seems that their approach also does not
have an obvious generalization to $D \ge 5$.
\end{enumerate}

\subsection{Topology theorems}
\label{sec:top}
We next summarize what is known about the topology of higher dimensional
black hole spacetimes, and what the differences to 4 dimensions are. Of interest are:

\begin{enumerate}
\item The topology of the event horizon cross sections, $B$.
\item The topology of the domain of outer communications, $\M$.
\end{enumerate}

\noindent
{\bf 1) Horizon Topology:} The topological (actually $C^\infty$-) invariant that is known
to characterize the topology of the horizon $B$ is the so-called ``Yamabe number''.
This is defined as follows. Let $\gamma$ be any Riemannian metric on $B$, not necessarily
that induced from the spacetime metric $g$. Define the number $Y[\gamma,B]$, depending on the conformal class of $\gamma$ by
\ben
Y[\gamma,B] := \inf_{\varphi>0, \varphi \in C^\infty}
               \frac{ \displaystyle\int_B R(\tilde \gamma) \ \D \tilde S
                    }{\left(
                            \displaystyle\int_B \D \tilde S
                      \right)^{\frac{D-4}{D-2}}} \ ,
\een
where $\tilde \gamma = \varphi^2 \gamma$ is the conformally transformed metric, and $\D \tilde S$ its integration
element. The Yamabe invariant is given by $Y[B] = \sup_{[\gamma]} Y[B,\gamma]$, where the sup is over all
conformal classes. In $D=4$, where $B$ is 2-dimensional, $Y[B]$ is up to a constant equal to the Euler characteristic of $B$.

Einstein's equations give constraints on the Yamabe invariant of horizon cross sections, and in fact also on more general $D-2$-dimensional compact surfaces $B \subset \M$ satisfying certain geometrical conditions.
These geometric conditions are that $B$ is a ``stably marginally outer trapped surface'' (MOTS). This
notion is defined as follows, see e.g.~\cite{marsanderson} for discussion.
We suppose that, on $B$, we have defined a pair of null vector fields $n,l$, orthogonal to $B$, such that $g(n,l) = 1$. $n$ is chosen future pointing, and is extended to generate a congruence of affine null geodesics, i.e. a null sheet $\mathscr N$. $l$ is
chosen ``outward pointing'', parallel transported along $\mathscr N$, and extended off $\N$ by demanding that it
be tangent to another congruence of null geodesics. $n,l$ are completely fixed once they
have been defined on $B$, and at each point $p$ of $B$, they are uniquely fixed
up to a rescaling keeping $g(n,l)=1$. Letting $\theta_{n,l}$ be the respective null
expansions, it is demanded that there exists a choice of $n,l$, such that
\ben
\theta_n = 0 \ , \quad \pounds_l \theta_n \ge 0 \ , \quad \text{on $B$} \ .
\een
The first condition states that the expansion on $\mathscr N$ in the $n$ direction vanishes, and the second one is satisfied if $\theta_n$ is non-negative
slightly outside of $\mathscr N$. An example of a MOTS is the event horizon $\H$ of a black hole (see~\cite{mars12} (proposition~3) and also~\cite{HW}),
see the following picture, so the following arguments in particular apply to that case.

\begin{center}
\begin{tikzpicture}

\node (rectnw) at (-6,2.3) {};
\node (rectne) at (8,2.3) {};
\draw[thick] (rectnw.center) -- (rectne.center);

%
\node (coneupperleft) at (-3.5,8) {};
\node (conemiddleleft) at (-3,3) {};
\node (ellipsel1) at (-2, 1.5){};
\node (coneintersL) at (-.8,.7) {};
\node (conebase) at (0,-.2) {};
\node (coneintersR) at (.8,.7) {};
\node (ellipser1) at (2, 1.5){};
\node (conemiddleright) at (3,3) {};
\node (coneupperright) at (3.5,8) {};
%
%


\draw[ opacity=0]
(coneupperleft.center) [rounded corners=15pt] -- node[pos=.15](n3end){} node[pos=.4](ellipsel3){} node[pos=.65](n2end){} node[pos=.9](ellipsel2){}
(conemiddleleft.center) [rounded corners=15pt]  -- (ellipsel1.south)[rounded corners=15pt]--
 (coneintersL.center)[rounded corners=1pt]  --   (conebase.center) [rounded corners=15pt] -- (coneintersR.center)[rounded corners=15pt]  -- (ellipser1.south)[rounded corners=15pt]
 --   node[pos=.5](HArrowtip){}
(conemiddleright.center)[rounded corners=15pt]  -- node[pos=.1](ellipser2){} node[pos=.6](ellipser3){}
(coneupperright.center);
%
\shadedraw[shading=axis, shading angle=90, fill opacity=.8] (ellipsel1)[rounded corners=15pt] -- (coneintersL.center)[rounded corners=1pt] -- (conebase.center)[rounded corners=15pt] -- (coneintersR.center) -- (ellipser1);

\shadedraw[left color=blue, right color=white, shading=axis, shading angle=90, fill opacity=.7] let \p1 = (coneintersL), \n1={veclen(\x1,0)} in (rectne.center) -- (5.5,\y1) node (rectse){} -- (-8.5,\y1) node (rectsw){} -- (rectnw.center) ;

\shadedraw[shading=axis, shading angle=90, fill opacity=.8] let
	\p1 = (ellipsel1),
	\n1 = {veclen(\x1,0)}
in  (coneupperleft.center)[rounded corners=15pt]   --
(conemiddleleft.center)[rounded corners=0pt]    -- (\x1,\y1) arc [x radius=\n1, y radius=.4cm, start angle=180, end angle=360][rounded corners=15pt]  --  (conemiddleright.center)[rounded corners=15pt] -- (coneupperright.center);

\draw[fill=white] (0,8) ellipse [x radius = 3.5, y radius= .5];
%
\draw[draw opacity = .4] (ellipsel1.center) to [ controls=++(90:.5) and ++(90:.5)]  (ellipser1.center);
\draw (ellipsel1.center) to [ controls=++(270:.5) and ++(270:.5)]  (ellipser1.center);
\draw[draw opacity = .4] (ellipsel2.center) to [ controls=++(90:.5) and ++(90:.5)]  (ellipser2.center);
\draw (ellipsel2.center) to [ controls=++(270:.5) and ++(270:.5)]  (ellipser2.center);
\draw[draw opacity = .4] (ellipsel3.center) to [ controls=++(90:.5) and ++(90:.5)]  (ellipser3.center);
\draw (ellipsel3.center) to [ controls=++(270:.5) and ++(270:.5)]  node[pos=.7, below](arrowBtip){} (ellipser3.center);
%
%

%
\node (cone1) at (-7.3,2) {};
\draw (cone1.center) -- ++(.5,-.5) node (cone1b){} -- ++(.5,.5);
\draw (cone1b) [yshift=.5cm] ellipse [ x radius= .5, y radius=.1];
\draw[->=latex,very thick] (cone1b.center) -- ++(0,1.5) node[right]{time} ;
\node (cone2) at (-7.3,4.5) {};
\draw (cone2.center) -- ++(.5,-.5) node (cone2b){} -- ++(.5,.5);
\draw (cone2b) [yshift=.5cm] ellipse [ x radius= .5, y radius=.1];
\draw[->=latex,very thick]  (cone2b.center) -- ++(0,1.5) node[right]{time} ;
\node (cone3) at (-7.3,7) {};
\draw (cone3.center) -- ++(.5,-.5) node (cone3b){} -- ++(.5,.5);
\draw (cone3b) [yshift=.5cm] ellipse [ x radius= .5, y radius=.1];
\draw[->=latex,very thick]  (cone3b.center) -- ++(0,1.5) node[right]{time} ;
\node (cone4) at (-5,2) {};
\draw[rotate=355] (cone4.center) -- ++(.5,-.5) node (cone4b){} -- ++(.5,.5);
\draw[rotate=355] (cone4b) [yshift=.5cm] ellipse [ x radius= .5, y radius=.1];
\draw[rotate=355,->=latex,very thick] (cone4b.center) -- ++(0,1.5);
\node (cone5) at (-5,4.5) {};
\draw[rotate=350] (cone5.center) -- ++(.5,-.5) node (cone5b){} -- ++(.5,.5);
\draw[rotate=350] (cone5b) [yshift=.5cm] ellipse [ x radius= .5, y radius=.1];
\draw[rotate=350,->=latex,very thick] (cone5b.center) -- ++(0,1.5);
\node (cone6) at (-5,7) {};
\draw[rotate=350] (cone6.center) -- ++(.5,-.5) node (cone6b){} -- ++(.5,.5);
\draw[rotate=350] (cone6b) [yshift=.5cm] ellipse [ x radius= .5, y radius=.1];
\draw[rotate=350,->=latex,very thick] (cone6b.center) -- ++(0,1.5);
\draw[->=latex, very thick] (ellipsel1.center) -- node[pos=.7] (n1){} ++(130:1cm) node[above] {$n$};
\draw[rotate=350] (n1) arc [x radius=.6cm, y radius=.1cm, start angle=180, end angle=310] ;
\draw[rotate=350] (n1) arc [x radius=.55cm, y radius=.1cm, start angle=160, end angle=10] ;
\draw[->=latex, very thick] (ellipsel1.center) -- ++(30:1cm) node[right]{$-l$};
\draw[->=latex,very thick] (ellipsel1.center) --  ++(77:1.2cm);
\draw[->=latex, very thick] (ellipsel2.center) -- node[pos=.6] (n2){} (n2end) node[left] {$n$};
\draw[->=latex, very thick] (ellipsel2.center) --  ++(5:1cm) node[right]{$-l$};
\draw[rotate=325] (n2) arc [x radius=.6cm, y radius=.1cm, start angle=180, end angle=300] ;
\draw[rotate=320] (n2) arc [x radius=.5cm, y radius=.1cm, start angle=180, end angle=0] ;
\draw[->=latex,very thick] (ellipsel2.center) --  ++(50:1.2cm);
\draw[->=latex, very thick] (ellipsel3.center) -- node[pos=.6] (n3){} (n3end) node[left] {$n$};
\draw[->=latex, very thick] (ellipsel3.center) -- ++(3:1cm) node[right]{$-l$};
\draw[rotate=325] (n3) arc [x radius=.6cm, y radius=.1cm, start angle=180, end angle=300] ;
\draw[rotate=320] (n3) arc [x radius=.5cm, y radius=.1cm, start angle=180, end angle=0] ;
\draw[->=latex,very thick] (ellipsel3.center) --  ++(50:1.2cm);
%
\node[below, scale=1.5] (sigma) at (rectse) {$\Sigma$};
\draw[latex-, very thick] (arrowBtip) -- ++(2, -.2) node[right]{$B=\{r=0, u=0\}$};
\draw[latex-, very thick] (HArrowtip.center) -- ++(1.5,1) node[right]{$\mathscr{H}=\{r=0\}$};
\node[below, scale=1.5] at (sigma.south) {$n=\frac{\partial}{\partial u}, l=\frac{\partial}{\partial r}$};
\end{tikzpicture}
\end{center}

A more pedestrian way of stating the MOTS condition is that, near $\mathscr N$, the metric $g$ takes Gaussian null form~\eqref{gaussian}, with $\mathscr N$ defined by $r=0$, $B$ defined by $u=0=r$, and with
$n=\partial/\partial u, l=\partial/\partial r$. (Since we take $u$ to be an affine parameter, the
metric coefficient of $\alpha$ in~\eqref{gaussian} is now in fact $r^2$ rather than only $r$.) The MOTS conditions are equivalent to $\gamma^{ab} \partial_u \gamma_{ab}=0$ and $\partial_r (\gamma^{ab} \partial_u \gamma_{ab}) \ge 0$. The theorem is~\cite{GallowaySchoen06,GA1}, see
also~\cite{HOY06} for antecedents:

\medskip
\noindent
\begin{thm}
(``Horizon topology theorem'')
If $B$ is a stably marginally outer trapped surface, and if the stress tensor satisfies the dominant
energy condition, then $B$ is of positive Yamabe type, $Y[B]>0$.
\end{thm}

A particularly simple proof of this theorem was given by~\cite{Racz08}. To combine the geometric conditions on
$B$, and the conditions on the stress tensor, i.e. in effect, the Einstein equations, one first derives from
the $ur$ and $ab$ components of the Einstein
equations in Gaussian null coordinates~\eqref{gaussian}:
\ben\label{Einstein}
\pounds_l \theta_n + \theta_n \theta_l = T_{ur} + \half [R(\gamma) + D^a \beta_a - \half \beta^a \beta_a] \ .
\een
Then, multiplying both sides by a testfunction $\varphi$, using the MOTS conditions, using $T_{ur} \ge 0$, applying the elementary inequality
\ben
\varphi^2 D^a \beta_a = D^a (\varphi^2 \beta_a) - 2\varphi(D^a \varphi) \beta_a \le D^a(\varphi^2 \beta_a) +
2 (D^a \varphi) D_a \varphi + \half \varphi^2 \beta^a \beta_a \ ,
\een
and integrating over $B$, one gets for $D \ge 5$:
\ben
\frac{ \displaystyle\int_B
       \left(
            4\frac{D-3}{D-4} (D^a \varphi) D_a \varphi + R(\gamma) \varphi^2
       \right) \ \D S
      }{
        \left(
              \displaystyle\int_B \varphi^{\frac{2(D-2)}{D-4}} \ \D S
        \right)^{\frac{D-4}{D-2}}
       }
\ge 0 \ .
\een
%
This inequality is known to imply $Y[B,\gamma] \ge 0$, hence $Y[B] \ge 0$ also. For $D=4$, one
simply integrates eq.~\eqref{Einstein} over $B$ and uses Gauss theorem to get rid of the total divergence
term. This immediately gives $\int_B R(\gamma) \ \D S \ge 0$, i.e. $B \cong S^2$ or $T^2$ by the Gauss-Bonnet-theorem. If one assumes the
{\em strict} MOTS condition ($\pounds_l \theta_n >0$ somewhere on $B$), one obtains the claim $Y[B] > 0$.
Actually, one can even show that $Y[B]>0$ without the strict MOTS
condition, by another argument~\cite{GallowaySchoen06,GA1}.

\medskip
\noindent
The condition that $Y[B]>0$ imposes a restriction on
the possible topologies of $B$. For example:
\begin{enumerate}
\item $D=4$: $Y[B]>0$ implies that the Euler characteristic of $B$ is positive, hence $B \cong S^2$.
Thus, one recovers Hawking's topology theorem~\cite{H72}.
\item $D=5$: In this case $B$ is a closed compact 3-manifold. It is known that $Y[B]>0$ implies that
$B$ is a connected sum
\ben
B \cong \#_i (S^3/\Gamma_i) \  \ \ \# k \cdot (S^2 \times S^1) \ ,
\een
where $\Gamma_i \subset O(4)$ are discrete subgroups (possibly trivial).
\item In dimensions $D>5$, the positive Yamabe condition becomes less and less restrictive.
\end{enumerate}

Further restrictions arise e.g. in $D=5$ in the case of stationary black holes, if
we combine these techniques with the topological censorship theorem (see below) and the rigidity theorem (see above). In that case, the rigidity theorem guarantees that the spacetime metric has a $U(1)$
symmetry in addition to being stationary. In particular, $B$ has an action of $U(1) \cong S^1$, and so is
a special case of a Seiffert 3-manifold. One can consider the factor space $\hat B = B/U(1)$, which
in general is not a manifold, but an ``orbifold'' (possibly with boundaries), characterized by singular points
with deficit angles $2\pi/p_i$, where $p_i \in \mathbb Z_+$. Then, using Einstein's equation, one derives that
the genus of $\hat B$ satisfies
\ben
\chi_{\rm orbifold}(\hat B) = 2-2 \ {\rm genus}(\hat B) - \sum_i \left( 1 - \frac{1}{p_i} \right) > 0 \ ,
\een
if $\hat B$ is closed, and one also derives constraints from the topological censorship theorem if
$\hat B$ has a boundary. The quantity on the left side is known as an invariant called ``orbifold Euler characteristic'' of the Seiffert fibration $B \to \hat B$.
For the case of positive orbifold Euler-characteristic, a complete classification of such fibrations is available in the mathematics literature, and this leads to~\cite{HHI10}:

\begin{thm} (``Refined topology theorem'')
For a stationary black hole in $D=5$
in a theory for which the topological censorship theorem and rigidity theorem hold, we must have
\ben
B \cong
\begin{cases}
S^3/\Gamma & \\
\# k \cdot (S^2 \times S^1) \ \ \ \#_i L(p_i, q_i) &
\end{cases}
\een
where the list of possible subgroups $\Gamma \subset O(4)$ is given in \cite{HHI10}, and where
each $L(p,q)$ is a lens space.
\end{thm}

The lens space corresponds to a particular quotient, but $\Gamma$
in the first line can be more general, and $S^3/\Gamma$ includes also e.g.
prism spaces, Poincare homology spheres etc.

Yet further restrictions arise in $D = 5$ and higher dimensions if one assumes that the spacetime
carries an isometric action not just of a single copy of $U(1)$, generic in view of
the rigidity theorem, but instead of multiple copies of $U(1)$, most likely non-generic.
The most stringent restrictions are obtained if the isometry group contains $U(1)^{D-3}$. Then it can be
shown that each horizon component must be one of the following~\cite{HollandsYazadjiev08b}:
\ben
B \cong \begin{cases}
S^3 \times \T^{D-5} \\
S^2 \times \T^{D-4} \\
L(p,q) \times \T^{D-5}
\end{cases}
\een
In particular, in $D=5$ we only have the possibilities $S^3, S^2 \times S^1, L(p,q)$. In particular,
more exotic quotients of $S^3$ other than the lens space are ruled out if the symmetry group contains
$U(1)^2$.

\medskip
\noindent
{\bf b) Domain of outer communication:} Much less is known about the possible topology of the domain of outer communication in general, except for static black holes (see sec.~\ref{sec:static}), where the topology of the Cauchy surface
$\Sigma$ connecting the horizon(s) and spatial infinity is flat space minus one ball per black hole horizon,
$\mr^{D-1} \setminus \cup_k B_k^{D-1}$. For stationary, but non-static, no comparable result is known, even though this seems to be the situation for most known black holes. For example, for Kerr the Cauchy surface $\Sigma$ also has topology $\Sigma \cong \mr^3 \setminus B^3$. The analogous statement is true for Myers-Perry black holes.
Similarly, for the black ring in 5 dimensions, $\Sigma \cong \mr^4 \setminus (S^1 \times B^2)$.

A simple general statement, which already follows from the requirement of global hyperbolicity
without even using Einstein's equations, is that
$\M \cong \mr \times \Sigma$. Another general
statement, which uses Einstein's equations, is the ``topological censorship theorem''~\cite{fsw,Galloway99,CGS09}, already alluded to above. In the most general form, it
can be stated as:

\begin{thm}
(``Topological censorship theorem'')
Let Einstein's equation hold with a stress tensor satisfying the null convergence condition
$T_{kk} \ge 0$ for any null vector $k$. If the spacetime is asymptotically flat or asymptotically Kaluza-Klein,
then any curve $c$ in $\M$ (or in $\Sigma$) with beginning and endpoint in the asymptotic region can be continuously deformed to
a curve $\tilde c$ that is entirely within the asymptotic region.
\end{thm}

In $D=4$, this theorem implies that $\Sigma$ is simply connected, and in particular furnishes an independent
proof that the horizon topology $B \cong S^2$: Indeed, suppose $B$ had handles. Consider a curve with endpoints at
infinity that is threaded through one of the handles of $B$. Such a curve clearly could not be smoothly deformed
to a curve that is entirely in the asymptotic region, a contradiction with the topological
censorship theorem.

In any dimension $D \ge 5$, the topological censorship theorem implies that $\M \cong \mr \times \Sigma$ (hence
$\Sigma$) is simply connected in the asymptotically flat case. This fact was already used in the proof of the
rigidity theorem, but it does not restrict the topology of $\Sigma$ very much in $D \ge 5$ dimensions. Tighter restrictions arise if we also use the $U(1)$ symmetry of the spacetime, which is guaranteed by the rigidity theorem.

\begin{enumerate}
\item As before, the restriction is most stringent in $D=5$, in which case $\Sigma$ is a simply connected,
asymptotically Euclidean, 4-manifold with inner boundary $B$. Then, using results from topology~\cite{orlik1,orlik2,fintushel}, it was shown in~\cite{HHI10} that
\begin{thm}
For a stationary asymptotically flat spacetime in a theory in which the
assumptions of the rigidity and topological censorship theorem are satisfied, we have
\ben
\Sigma \cong [\# m \cdot (S^2 \times S^2) \ \ \#  \ m' \cdot \mc P^2 \ \ \# \ m'' \cdot \overline{\mc P^2} \ \ \# \ \mr^4] \setminus {\rm BH}
\een
where $BH$ is a compact 4-manifold with boundary $B$.
\end{thm}
If the spacetime is spin, then the complex projective
spaces are absent in the decomposition. In all known solutions, the handles $S^2 \times S^2$ are also absent.
This result should maybe be contrasted with the decomposition of general 4-manifolds. As is well-known,
with any compact oriented 4-manifold $X$ one can associate a canonical pairing $Q_X: H_2(X, \mz) \times H_2(X, \mz) \to \mr$. That pairing associates with a pair of 2-cycles $C, C'$
the intersection number $\#(C \cap C')$, where each intersection point is counted
with a $\pm$ sign determined by the relative orientations. $Q_X$ defines a quadratic form
with entries in $\mz$. A deep theorem in topology~\cite{freed} states that $Q_X$
can be decomposed up to a similarity transformation as
\ben
Q_X =  m \cdot \left(
\begin{matrix}
0 & 1\\
1 & 0
\end{matrix} \right)  \oplus I_{m'} \oplus (-I_{m''}) \oplus m''' \cdot E_8
\een\label{qform}
for some $m, \dots, m''' \in \mz_+$, where $E_8$ is the Cartan matrix of the
exceptional Lie algebra with the same name, and $I_m$ the $m$-dimensional identity matrix.
Furthermore, $X$ is determined
topologically by its $Q_X$ uniquely. If the handle decomposition of $X$
only has $S^2 \times S^2$'s (corresponding to $m$) and complex projective spaces
$\mc P^2$ (corresponding to $m'$) and $\overline{\mc P^2}$ (corresponding to
$m''$), then one can see~\cite{orlik1, orlik2} that $m'''=0$, and the $E_8$-factors are absent. (If the manifold is spin, the
the complex projective spaces are also absent, $m'=0=m''$.) This result cannot be
applied directly to $\Sigma$, because it is not a compact manifold. However,
we can glue in a ball at infinity, and a suitable manifold at the horizon (see~\cite{HollandsYazadjiev08b} for details) to obtain from $\Sigma$ a compact
4-manifold in a canonical way, which admits again an action of $U(1)^2$. If we call
by abuse of notation the quadratic form of this manifold by $Q_\Sigma$, then we have:

\begin{thm}
For a rotating stationary black hole $\M \cong \mr \times \Sigma$, in $D=5$, the decomposition of $Q_\Sigma$
does not contain $E_8$'s.
\end{thm}

\item In higher dimensions $D \ge 6$, the topological censorship theorem
combined with the $U(1)$ symmetry from the rigidity theorem, does not seem to give
a lot of restriction. However, assume that we even have an action of $U(1)^{D-3}$. We {\em conjecture}:

\begin{conjecture}
Assume that $(\M,g)$ is spin, asymptotically flat or KK, has a Ricci-tensor
 satisfying the null-convergence condition, and admits an action of $U(1)^{D-3}$. Then the Cauchy
surface $\Sigma$ can be decomposed as
\ben
\Sigma \cong [\#_{k=2}^{D-3} \ m_k \cdot (S^k \times S^{D-1-k}) \ \ \# ({\rm asymptotic \ \ region})]
\setminus {\rm BH} \ ,
\een
where the asymptotic region depends on the precise boundary conditions; e.g. in the standard
KK setup $\mr^3 \times \T^{D-4}$.
\end{conjecture}

\end{enumerate}

\subsection{Weyl-Papapetrou coordinates and orbit space analysis}\label{ref:wp}

An important step in the uniqueness proof of the Kerr-Newman metric in $D=4$ is to bring the metric into the Weyl-Papapetrou
form. In 4 dimensions, the Weyl-Papapetrou form follows from the existence of the commuting Killing fields
$t,\psi$ and certain regularity assumptions about the global causal structure alone, whereas the
existence of $\psi$ in turn follows form the rigidity theorem, and $t$ is assumed to start with. Thus, the Weyl-Papapetrou form of the metric
for stationary black essentially (i.e., modulo the technical assumptions described in sec.~\ref{sec1}) does not imply any loss of generality for stationary
metrics.  However, in $D \ge 5$, the Weyl-Papapetrou form only seems to follow if one makes
the assumption that the number $N$ of rotational Killing fields is $N=D-3$. The rigidity theorem
only guarantees $N \ge 1$ rotational Killing fields, and this might well be the generic case.
Thus, it is likely that the Weyl-Papapetrou form
is not generic in higher dimensions, even in vacuum Einstein-gravity. Nevertheless, if one does
assume the existence of $D-3$ commuting axial Killing fields, (i.e. an isometric action of $U(1)^{D-3}$
on $(\M,g)$), then the generalized Weyl-Papapetrou can be shown to follow:

\begin{thm}\label{wpthm}
(``Weyl-Papapetrou-form'')
If the spacetime is asymptotically Kaluza-Klein with $3,4,5$ asymptotically flat
(``large'') dimensions, satisfies the vacuum Einstein equations,
if there are commuting Killing fields $t, \psi_1, \dots, \psi_{D-3}$ generating the isometry
group $G = \mr \times U(1)^{D-3}$, and if the spacetime satisfies satisfies a causal
regularity condition mentioned in sec.~\ref{sec1}, then the metric can be brought into Weyl-Papapetrou form
\ben\label{wp}
g = - \frac{r^2 \, \D \tau^2}{\det f} + \e^{-\nu} (\D r^2 + \D z^2)
+
f_{ij}( \D \varphi^i + w^i \, \D \tau)(\D \varphi^j + w^j \, \D \tau) \, ,
\een
globally, but away from the horizon $\H$ and any axis of rotation.
\end{thm}
The notation for the coordinates in the Weyl-Papapetrou form is analogous to
that given before in 4 dimensions: $\varphi^i$ are $2\pi$-periodic coordinates such that the rotational Killing fields $\psi_1, \dots, \psi_{D-3}$
take the form $\psi_i=\partial/\partial \varphi^i$, and $\tau$ is a coordinate such that
the timelike Killing field takes the form $t = \partial/\partial \tau$. In other words,
the metric functions $f_{ij}, w^i, \nu$ are independent of $\tau, \varphi^1, \dots, \varphi^{D-3}$,
and only depend on $z \in \mr, r > 0$. The argument establishing~\eqref{wp} is outlined below.

It is important to emphasize that the Weyl-Papapetrou form does not reflect in a very transparent way either the global nature of $\M$,
nor the global nature of the action of the symmetry group on $\M$ in $D \ge 5$. Both have to be taken into account
and understood to appreciate the meaning of the coordinates, and to actually derive the form of the metric.
Broadly speaking, the points labeled by $r=0$ describe points in $\M$ which are either (i) on the horizon $\H$, or (ii) lie on an ``axis of rotation''. By the latter one means in higher dimensions points such that a non-trivial
linear combination of the rotational Killing fields $\psi_i$ vanishes. In case (i), this follows from the fact
that $t = \partial/\partial \tau$ has zero norm on $\H$. For (ii), this can be seen from the fact that, for such
points $f_{ij}$ fails to have full rank, hence $\det f = 0$, hence from the first term in $g$, $r=0$.
This will be described in more detail when we come to the proof of~\eqref{wp}.
In particular, we will describe how non-trivial topologies of $\M$ are compatible with~\eqref{wp}. We also note that
the relation between $r,z$, and the asymptotically Cartesian (spatial) coordinates $x_1, \dots, x_s$ in the asymptotically
KK-region~(see sec.~\ref{sec1}) is
\ben
(r,z) \sim \begin{cases}
(\sqrt{x_1^2 + x_2^2}, x_3) & \text{if $s=3$}\\
(\sqrt{(x_1^2+x_2^2)(x_3^2+x_4^2)}, \half(x_1^2 + x_2^2 - x_3^2 - x_4^2)) & \text{if $s=4$}
\end{cases}
\een
The theorem~\ref{wpthm} can be proved also for more general Einstein-matter systems, see sec.~\ref{sec:other}. The proof of
this result is rather more non-trivial than in $D=4$, due to new global considerations about the topology of $\M$
and the nature of the action of the isometry group $G=\mr \times U(1)^{D-3}$. Therefore we will outline it here,
for details see~\cite{HollandsYazadjiev08b}. For definiteness, we will now outline the proof in the case of vacuum general relativity.

Because the symmetry group has $(D-2)$ dimensions, the metric will, in a sense, depend non-trivially only on two remaining coordinates, and the Einstein equations will consequently reduce to a coupled system of PDE's in these variables ($r,z$ above, but we need to explain how exactly we choose them!). However, before one can study these equations, one must understand more precisely the nature of the two remaining coordinates, or, mathematically speaking, the nature of the orbit space $\M/[\mr \times U(1)^{D-3}]$. The quotient by $\mr$ simply gets rid of a global time coordinate, so one is left with the quotient of a spatial slice $\Sigma$ by $U(1)^{D-3}$. To get an idea about the topological properties of this quotient, we consider the following two simple, characteristic examples in the case $\dim \Sigma = 4$, i.e. $D=5$.

 The first example is $\Sigma = \mr^4$, with one factor of $U(1) \times U(1)$ acting by rotations in the 12-plane and the other in the 34-plane. Introducing polar coordinates $(R_1,\varphi_1)$ and $(R_2,\varphi_2)$ in each of these planes, the group shifts $\varphi_1$ resp. $\varphi_2$, and the quotient is thus given simply by the first quadrant $R_1>0, R_2>0$ , which is a 2-manifold with two boundary components, i.e. the semi-axis, and the corner where the two axis meet. The first boundary component corresponds to places in $\mr^4$ where the Killing field $\psi_1 = \partial/\partial \varphi_1$ vanishes, whereas the second boundary component to places where $\psi_2 = \partial/\partial \varphi_2$ vanishes. On the corner, both Killing fields vanish and the group action has a
 fixed point. The second example is the cartesian product of a plane with a 2-torus, $\Sigma = \mr^2 \times \T^2$. Letting
 $(x_1,x_2)$ be cartesian coordinates on the plane, and $(\varphi_1, \varphi_2)$ angles on the torus, suppose the group action is generated
 by the vector fields $\psi_1 = \partial/\partial \varphi_1$ and by $\psi_2 = \alpha \partial/\partial \varphi_2 + \beta
(x_1 \partial/\partial x_2 - x_2 \partial/\partial x_1)$, where $\alpha,\beta$ are integers.
These vector fields do not vanish anywhere, but there are discrete group elements leaving certain points invariant. The quotient is now a cone with deficit angle $2\pi/\alpha$.

\begin{figure}[h]
  \begin{center}
  \includegraphics[width=15cm]{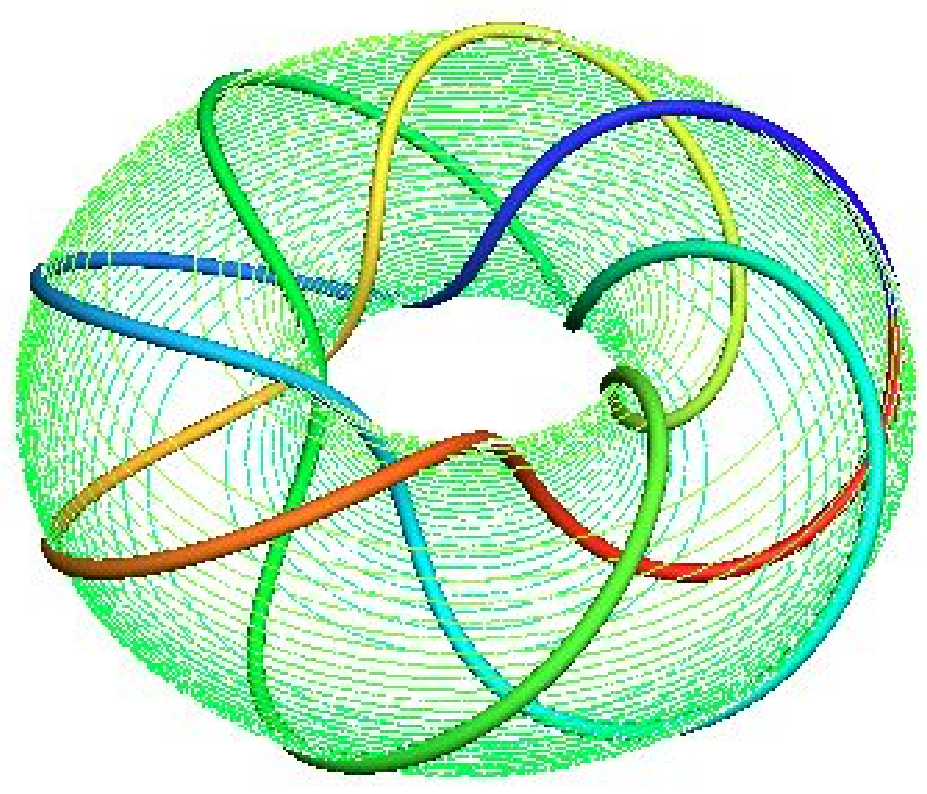} \hspace{1cm}
  \caption{\label{fig1}
   \small{The numbers $(p,q)$ may be viewed as winding numbers associated with the generators of the 2-torus
   generated by the two axial Killing fields. In such a torus, an $U(1)$-orbit winds around the first $S^1$ generator $n$-times as it goes $p$-times
around the other $S^1$-direction. Here ($qn \equiv 1 \, {\rm mod} \ p$). The figure shows the situation for $p=3$, $n=7$.} }
  \end{center}
 \end{figure}

The general case turns out to be locally the same as in these examples~\cite{orlik1,orlik2} ($D=5$) \cite{HollandsYazadjiev08} (general $D$). In fact, one can show that the quotient $\Sigma/[U(1) \times U(1)]$ is a 2-dimensional conifold with boundaries and corners. Each boundary segment is characterized by a different pair $(p,q)$ of integers such that $p \psi_1 + q \psi_2 = 0$ at corresponding points of $\Sigma$, see fig.~\ref{fig1}.

For subsequent boundary segments adjacent on a corner labeled by $\underline{v}_i = (p_i,q_i)$ and $\underline{v}_{i+1}=(p_{i+1},q_{i+1})$, we have the condition
\ben\label{constr}
\det \left(
\begin{matrix}
p_i & p_{i+1}\\
q_i & q_{i+1}
\end{matrix}
\right) = \pm 1 \, .
\een
Each conical singularity is  characterized by a deficit angle, i.e. another integer. In higher dimensions, there is a similar result; now a boundary segment is e.g. characterized by $(D-3)$-tuple of integers
$\underline v_i = (v_{1i}, \dots, v_{(D-3)i})$ (``winding numbers''), and the compatibility condition at the corners is somewhat more complicated, see the theorem below.

In the case where $\Sigma$ is the spatial section of a black hole spacetime, there are further constraints coming from
Einstein's equations, and the orientability of the spacetime. The topological censorship theorem is seen to imply, using various methods in algebraic topology, that the 2-dimensional orbit space $\hat \Sigma = \Sigma/U(1)^{D-3}$ {\em cannot have any conifold points, nor holes, nor handles,} and therefore has to be diffeomorphic to an upper half plane $\hat \Sigma \cong \{(r,z) \mid r>0\}$. The boundary segments correspond to intervals on the boundary $(r=0)$ of this upper half plane and are places (``axis'') in the manifold $M$ where a linear combination of the rotational Killing fields vanish, or to a horizon. Here the Killing fields do not vanish except where an ``axis'' meets a horizon. Furthermore, with each boundary segment $(z_J, z_{J+1})$, one can associate its length $l_J \ge 0$ via the metric $g$ (see below).

\begin{center}
\begin{tikzpicture}[scale=1.1, transform shape]
\shade[left color=gray] (-4,0) -- (5,0) -- (5,3)  -- (-4,3) -- (-4,0);
\draw[very thick,color=red] (1,0) -- (3,0);
\draw[very thick,color=red,dashed] (3,0) -- (4,0);
\draw[->,very thick,color=red] (4,0) -- (5,0);
\draw (5,0) node[right,color=red]{$z$};
\draw[very thick,color=red] (-4,0) -- (0,0);
\draw[very thick,color=blue] (0,0) -- (1,0);
\draw (.5,0) node[below,color=blue]{$\hat \H = \H/G$};
\draw (1.6,2.8) node[below]{$\hat \M=\M/G = \{(r,z) \mid r>0\}$};
\filldraw[color=blue] (1,0) circle (.05cm);
\filldraw[color=blue] (0,0) circle (.05cm);
\filldraw[color=red] (-1.8,0) circle (.05cm);
\draw[color=red] (-1.8,0) node[below]{$z_{J-1}$};
\filldraw[color=red] (-3.7,0) circle (.05cm);
\draw[color=red] (-3.7,0) node[below]{$z_{J}$};
\filldraw[color=red] (-1,0) circle (.05cm);
\draw[black] (4.5,0) node[below]{$\underline{v}_{0}$};
\filldraw[color=red] (2,0) circle (.05cm);
\filldraw[color=red] (2.5,0) circle (.05cm);
\draw (2.5,0) node[color=red,below]{$z_0$};
\draw (-2.8,0) node[below]{$\underline{v}_J$};
\end{tikzpicture}
\end{center}

Thus, in summary, one has~\cite{HollandsYazadjiev08b}:

\begin{thm}\label{thm:orbit}
(``Orbit space theorem'') Let $(\M, g)$ be asymptotically flat or KK, with Ricci
tensor satisfying the null-convergence condition. If one assumes the isometry group $G = \mr \times U(1)^{D-3}$, then the orbit space
$\hat \M = \M/G$ is homeomorphic to an upper half plane $\{(r,z) \mid r>0\}$. Furthermore, the
boundary $r=0$ can be thought of as divided up into a collection of intervals $(-\infty, z_1), (z_1, z_2),
\dots, (z_n, +\infty)$, each of which either represents the orbit space $\hat \H = \H/G$ of the horizon
(one interval per horizon component, if multiple horizons are present), or an axis in the spacetime
where a linear combination $\sum_i v_{iJ} \psi_i$ of the rotational Killing fields vanishes. The quantity $\underline{v}_J
= (v_{1J}, \dots, v_{(D-3) J}) \in \mz^{D-3}$ is a vector associated with the $J$-th interval which necessarily has
integer entries\footnote{It is unique up to sign if we impose, as we will, that $g.c.d.(v_{1J}, \dots, v_{(D-3) J})=1$.}.
For adjacent intervals $J$ and $J+1$ (not including the horizon), there is a compatibility condition stating that
the collection of minors $Q_{kl} \in \mz, \,\, 1 \le k < l \le D-3$ given by
\ben
Q_{kl} = \left|\det \left(
\begin{matrix}
v_{k(J+1)} & v_{kJ}\\
v_{l(J+1)} & v_{lJ}
\end{matrix}
\right) \right| \,
\een
have greatest common divisor ${g.c.d.}\{ Q_{kl} \}= 1$
\end{thm}

Some examples in $D=5$ are summarized in the following tables
\begin{center}
\begin{tabular}{|c|c|c|c|}
\hline
               & Interval Lengths & Vectors (Labels) & Horizon \\
\hline
Myers-Perry & $\infty, l_1, \infty$ & $(1,0),
(0,0), (0,1)$ & $S^3$ \\
\hline
Black Ring & $\infty, l_1, l_2, \infty$ & $(1,0),
(0,0), (1,0), (0,1)$ & $S^2 \times S^1$\\
\hline
Black Saturn & $\infty, l_1, l_2, l_3, \infty$ & $(1,0), (0,0), (0, 1), (0,0), (0,1)$ & $S^3 \ \
{\rm and} \ \ S^2 \times S^1$\\
\hline
Black String & $\infty, l_1, \infty$ & $(1,0), (0,0), (1,0)$ & $S^1 \times S^2$\\
\hline
Black Di-Ring & $\infty, l_1, l_2, l_3, l_4, \infty$ & $(1,0),(0,0),(1,0),(0,0),(1,0),(0,1)$ & $2 \cdot (S^1 \times S^2) $  \\
\hline
Orth. Di-Ring & $\infty, l_1, l_2, l_3, l_4, \infty$ & $(1,0),(0,0),(1,0),(0,1),(0,0),(0,1)$ & $2 \cdot (S^1 \times S^2)$  \\
\hline
Minkowski & $\infty, \infty$ & $(1,0), (0,1)$ & ---\\
\hline
\end{tabular}
\end{center}
The explicit form of the metric may be found in~\cite{MP86} (Myers-Perry),
\cite{PS06,ER02b} (Black ring), \cite{Elvang&Figueras07} (Saturn),
\cite{IguchiMishima07} (Di-Ring), and~\cite{Izumi} (Orthogonal Di-Ring).
The following interval structure would represent a ``Black Lens'',
respectively an exotic spherical black hole embedded in
an ambient space containing a factor of $\mc P^2$, if
such solutions would exist:
\begin{center}
\begin{tabular}{|c|c|c|c|}
\hline
& Interval Lengths & Vectors (Labels) & Horizon Topology \\
\hline
Black Lens & $\infty,l_1, l_2, \infty$ & $(1,0),
(0,0), (1,p), (0,1)$ & $L(p,1)$\\
\hline
Exotic MP & $\infty,l_1, l_2, l_3, l_4,\infty$ & $(1,0),
(0,0), (0,1), (1,1), (1,0), (0,1) $ & $S^3$\\
\hline
\end{tabular}
\end{center}
In these tables, the interval $(0,0)$ corresponds to a horizon. If $\underline{v}_{h-1}$
resp. $\underline{v}_{h+1}$ represent the vectors adjacent to a horizon interval $(z_h, z_{h+1})$
then the parameter $p = |{\rm det} \, (\underline{v}_{h-1}, \underline{v}_{h+1})|$ is related to the
different horizon topologies by:

\begin{figure}
\begin{center}
\begin{tabular}{c|c}
$p$ & Topology of $\H$ \\
\hline
$0$ & $S^2 \times S^1$ \\
$\pm 1$ & $S^3$ \\
other & $L(p,q)$
\end{tabular}
\caption{\label{table1}
   \small{The invariant $p= |{\rm det} \, (\underline{v}_{h-1}, \underline{v}_{h+1})|$ characterizes the
   different horizon topologies.} }
\end{center}
\end{figure}

Note also that the first and last vector $\underline{v}_0, \underline{v}_N$ in the above solutions
is always $(1,0)$ resp. $(0,1)$. This corresponds to the fact that these 5-dimensional solutions
are asymptotically flat in all 5 directions. For 5-dimensional
solutions which are asymptotically Kaluza-Klein with one compactified extra dimension, the
first and last vectors would be equal, e.g. $(0,1)$ and $(0,1)$. In general, the relationship between
$\underline{v}_0, \underline{v}_N$ and the asymptotic conditions is as follows. First we form
the parameter $p = |{\rm det} \, (\underline{v}_{0}, \underline{v}_{N})|$. Then

\begin{figure}
\begin{center}
\begin{tabular}{c|c}
$p$ & Asymptotic conditions \\
\hline
$0$ & $\mr^{3,1} \times \T$ (Kaluza-Klein)\\
$\pm 1$ & $\mr^{4,1}$ (Minkowskian) \\
other & $\mr^{4,1}/\mz_p$ (locally Minkowskian)
\end{tabular}
\caption{\label{table2}
   \small{The invariant $p= |{\rm det} \, (\underline{v}_{0}, \underline{v}_{N})|$ characterizes the
   different possible asymptotic behaviors.} }
\end{center}
\end{figure}

The last item in the table means that the 4-dimensional spatial slice in the asymptotic region is locally
asymptotically Euclidean, i.e. a large sphere $S^3$ near spatial infinity
is replaced by the quotient $S^3/\mz_p$ of $S^3$ by a discrete cyclic group.
Trivial examples of such spacetimes are  $g = -\D \tau^2 + \D s^2_{\rm Instanton}$, where
the spatial part is a suitable 4-dimensional gravitational instanton. These static spacetimes of course do not
contain a black hole, but black hole spacetimes of this nature have been constructed by~\cite{teo}.

The numbers $\{l_J\}$ and the
assignment of the labels
$\{\underline{v}_J\}$ were also considered from a local perspective by~\cite{Harmark-Olesen05}, \cite{Harmark04}. However,
we note that, in these references, neither the condition that the components be integers,
nor the determinant conditions for adjacent interval vectors and
their relation to the horizon topology
were obtained. Furthermore, the interval vectors considered in~\cite{Harmark04} have
$D-2$ components, rather than $D-3$.

\medskip
\noindent
So far our considerations have been essentially topological; in particular we have not said how exactly the
coordinates $(r,z)$ on $\hat \M = \M/G$ are to be chosen relative to the metric $g$. The key trick, generalized from~\cite{Carter71}, is now to make a special choice
using the Einstein equations. First, $r$ is defined by
\ben\label{rdef}
r^2 = - \det \left(
\begin{matrix}
g(t,t) & g(t, \psi_i) \\
g(\psi_i, t) & g(\psi_i, \psi_j)
\end{matrix}
\right)
\een
That the right side is positive, or equivalently, that the span of $t,\psi_1, \dots, \psi_{D-3}$ is
a {\em timelike} subspace in each tangent space $T_x \M$, is actually rather non-obvious, and
global arguments are required to demonstrate it~\cite{CC08}. However, it then immediately follows
that $\hat \M$ inherits a {\em Riemannian} metric, $\D \hat s^2_2$, and it furthermore follows from the vacuum
Einstein equations that $r$ is a harmonic function on $\hat \M$ w.r.t. this Riemannian metric. It can be seen from this fact, together with an application of the maximum principle and the uniformization theorem~\cite{CC08}, that $r$ is a well-defined coordinate on $\hat M$, i.e.
that its gradient cannot vanish. We may supplement $r$ by a second, globally defined, conjugate harmonic coordinate $z$ on $\hat \M$
\ben
\D r = \star_2 \D z \ .
\een
Together, $(r,z)$ then provide the desired geometrically preferred coordinate system on $\hat \M =
\{ (r,z) \mid r>0 \}$. The interval lengths $l_J$ are defined by $l_J = z_{J+1}-z_J$ in this special
coordinate system. The induced metric on $\hat \M$ can be written by construction as $\D \hat s_2^2 = \e^{-\nu(r,z)}(\D z^2 + \D r^2)$ for some function $\nu$, because $(r,z)$ is a conjugate pair.

The $(r,z)$ coordinates can then complemented by coordinates $\tau \in \mr$, and
$2\pi$-periodic coordinates $\varphi^1, \dots, \varphi^{D-3}$ in such a way that the Killing fields
are $t = \partial/\partial \tau, \psi_i = \partial/\partial \varphi^i$.
The final step is to show that the family of subspaces
${\rm span}(t,\psi_1, \dots, \psi_{D-3})^\perp \subset T\M$ is integrable. This is shown
via Frobenius' theorem, in combination with Einstein's equation. In other words, we see in this way that
the metric does not have any cross terms between $(\tau, \varphi^1, \dots, \varphi^{D-3})$, respectively
$(r,z)$, and the Weyl-Papapetrou form of the metric~\eqref{wp} then follows.

In view of generalizations to more general systems of equations, it is maybe useful to show exactly how
integrability is proven. By the ``differential form version'' of Frobenius' theorem, we need to show that
$\D \xi_I = \sum_J \alpha_{IJ} \wedge \xi_J$, where $\xi_I$ collectively denotes the $D-2$ Killing fields, identified with 1-forms using the metric. This is equivalent to
\bena
\xi_1 \wedge \cdots \wedge \xi_{D-2} \wedge \D \xi_J = 0 \quad
\text{for all $J$.}
\eena
Taking a Hodge-dual of this equation and applying $\D$, gives, using Einstein equations, standard
identities for Killing vectors, and the commuting nature of the Killing fields,
\ben\label{killid}
\D \star_D (\xi_1 \wedge \cdots \wedge \xi_{D-2} \wedge \D \xi_J) = -8\pi \star_D [\xi_1 \wedge \cdots \wedge \xi_{D-2}
\wedge T(\xi_J)]
\een
where $T(\xi_J)$ is the 1-form obtained by dotting the vector $\xi_J$ into one of the indices of the stress
tensor. The right hand side of course vanishes automatically in the vacuum, but also e.g. for the
class of theories described by~\eqref{genth} (assuming all the matter fields are Lie-derived by
the $\xi_I$). Thus, we see that $\star_D (\xi_1 \wedge \cdots \wedge \xi_{D-2} \wedge \D \xi_J)$ is
a constant. But we also know from the above analysis of the orbit space that at least one linear
combination of the $\xi_I$ must vanish somewhere on $\M$, hence it is zero, so we are done.

\subsection{Non-linear sigma-model reduction}\label{sec:sigma}

The Weyl-Papapetrou form considerably constrains the metric, but it is still not well-suited either
to the proof of uniqueness, or the generation of new solutions from old ones. For this, yet another set
of ideas is required, namely the embedding of dimensionally reduced gravity models into certain non-linear
sigma-models. (The oldest manifestation of this fundamental idea is the discovery by Ehlers~\cite{Ehlers}
of a ``hidden'' $SL(2)$ symmetry of the stationary axisymmetric Einstein equations.) This embedding is best described if we forget, again, for the moment about the asymptotically
time-like Killing vector. Then the metric~\eqref{wp} can be written in standard Kaluza-Klein form,
\ben\label{kkansatz}
g = (\det f)^{-1} \ g_3 + f_{ij}(\D \varphi^i + A^i)(\D \varphi^j + A^j) \ ,
\een
where $g_3$ is the induced line-element on the orbit space and time, $\hat \M \times \mr$
(parameterized by the coordinates $(\tau, r, z)$), and where each $A^i$ is a one-form on this space. Explicitly,
\ben
g_3 = - r^2 \, \D \tau^2 + \det f \ \e^{-\nu} (\D r^2 + \D z^2) \ , \quad
A^i=w^i \ \D \tau \ .
\een
One now takes this metric and plugs it into the field equations; for simplicity we will focus on the
vacuum case, but other cases will be discussed below in sec.~\ref{sec:other}. This gives the
``Maxwell equation''
\ben
\D \bigg( \det f \cdot f_{ij} \star_3  \D A^j \bigg) = 0 \ ,
\een
implying locally the existence of potentials $\chi_i$ on $\mr \times \hat \M$ satisfying
\ben\label{chieq}
\D \chi_i = 2 \ \det f \cdot f_{ij}  \star_3  \D A^j \ \ .
\een
These so-called ``twist potentials'' are in fact defined globally~\cite{HollandsYazadjiev08b}. Following Maison~\cite{Maison79},
one next defines the matrix valued function $\Phi$ by
\ben\label{phidef}
\Phi = \left(
\begin{matrix}
(\det f)^{-1} & -(\det f)^{-1} \chi_i \\
-(\det f)^{-1} \chi_i & f_{ij} + (\det f)^{-1} \chi_i \chi_j
\end{matrix}
\right) \quad .
\een
The matrix $\Phi$ is symmetric, positive definite, and $\det \Phi = 1$. It turns out
that the vacuum field equations are equivalent to a coupled system of equations for
$g_3=\D s^2_3$ and the matrix field $\Phi$, which are precisely the Euler-Lagrange equations of the
action
\ben\label{Sreduced}
I = \int \half R_3 \ \star_3 1 + \frac{1}{8} {\rm Tr} \Big( \Phi^{-1} \D \Phi \wedge \star_3 \Phi^{-1} \D \Phi \Big) \ .
\een
Here, $R_3$ is the Ricci-scalar for the 3-dimensional Lorentzian metric $g_3$ on $\mr \times \hat \M$.
This is the action of a 3-dimensional gravitating sigma-model. Since $\Phi$ is a real, unimodular, symmetric,
positive definite $(D-2)$-dimensional matrix, it can be written as $\Phi = S^T S$ for some real matrix $S$
of determinant $1$, i.e. an $SL(D-2)$ matrix. This matrix is defined up to $S \to RS$ for some rotation
$R \in SO(D-2)$, so in this sense, $\Phi$ may be thought of as taking its values in the coset manifold
$X = SL(D-2)/SO(D-2)$. The trace term in the action is precisely the standard kinetic term
for the standard Riemannian metric $\mathcal G$ on $X$.

The equations of motion for the $\Phi$-field are equivalent to the conservation laws of the
$\dim SL(D-2)$ currents ${\mathcal J}^I = {\rm Tr} \ (T^I \Phi^{-1} \D \Phi)$, where $T^I$
are the generators of the Lie algebra $\frak{sl}(D-2)$. We may at this stage remember that
we have another Killing vector, $t$, or equivalently, that the metric $g_3$ does not depend
on the coordinate $\tau$. Using the explicit form of this metric then allows one to write the
equation of motion for the matrix field $\Phi$ as an equation on $\hat \M$, with metric
$\D \hat s^2_2 = \e^{-\nu}(\D r^2 + \D z^2)$, i.e. an equation in $r,z$ alone. Explicitly, this is
\ben\label{ref}
\frac{1}{r} \D (r\star_2 {\mathcal J}^I) = 0 \quad
\Longleftrightarrow \quad
\partial_z (\Phi^{-1} \partial_z \Phi) + \frac{1}{r} \partial_r (r \Phi^{-1} \partial_r \Phi) = 0 \ .
\een
The remaining Einstein equations for the sigma model action $I$ for the metric $g_3$ then give
an equation for $\nu$. Since the equations for the currents are independent of $\nu$, it follows
that the sigma-model equations are decoupled from this equation. The two sets of equations can hence be solved successively. Finally,
once we have $\nu, \Phi$, we can get $A^j = w^j \ \D \tau$ by inverting eq.~\eqref{chieq}. In this
way, we can in principle solve for all unknown functions in the Weyl-Papapetrou form~\eqref{wp}.

The sigma-model formulation of the Einstein equations has several important uses. For us, the
most important one is that one can obtain ``divergence identities'' that play a key role
in the proof of the uniqueness theorem, see the next subsection. However, another important consequence of the
sigma model formulation is the possibility to generate new solutions from old ones.
A simple but powerful way to do this is to
apply to a $\Phi$ representing a given solution a similarity transformation $R \Phi R^{-1}$, where $R$ is
any constant rotation matrix. Since such a transformation is a symmetry of the action, we get a new
solution, which, because the $f_{ij}$ and $\chi_i$ components of $\Phi$ are mixed, will in general differ
non-trivially from the original one. This type of transformation gives a group action of $SO(D-2)$ on the space of solutions. A less obvious fact is that this group can be enhanced to to the infinite-dimensional ``Geroch-group''. For 4-dimensions, a the sigma-model perspective on this construction~\cite{geroch} may be found in~\cite{maison}, but the same arguments should go through in $D$-dimensional vacuum gravity as well. A related feature of the
equations~\eqref{ref} is that solutions may be generated by the same sort of techniques (``inverse scattering method'')
developed originally for the Sine-Gordon- and KdV-equations. For a detailed explanation how to
apply such techniques, we refer to the textbook~\cite{verdaguer}. While such techniques
have been widely and successfully applied to obtain highly non-trivial higher dimensional metrics, an unpleasant feature of
 all solution generating techniques is that, starting from a given regular, say asymptotically flat, stationary black hole solution, the newly generated solution may neither be regular, nor asymptotically flat, nor even
describing a black hole.

To see more clearly the relationship between regularity and the global considerations about the orbit space made
in the previous subsection, in particular the ``interval structure'', suppose we have a given single black hole solution characterized by a general interval structure:

\begin{center}
\begin{tabular}{|c|c|}
\hline
Interval Lengths & Vectors (Labels) \\
\hline
$\infty,l_1, \dots,l_h,\dots, l_N,\infty$ & $\underline{v}_0, \underline{v}_1,
\dots, \underline{0}, \dots, \underline{v}_N, \underline{v}_{N+1} $ \\
\hline
\end{tabular}
\end{center}
Here, $l_h$ corresponds to the horizon interval $(z_h,z_{h+1})$ on the
boundary of $(r,z)$-space. Now, recall that the integer vectors $\underline{v}_J$
tell us which linear combination $\sum v_{Ji}\psi_i=0$ vanishes at the axis
represented by the respective interval. Hence, we must have at the $J$-th interval
$z \in (z_J, z_{J+1})$:
\ben\label{bndy1}
\sum_i f_{ij}(r=0,z) v_{iJ} = 0 \ .
\een
Also, one can see that the potentials $\chi_i$ are constant along the boundary $r=0$,
except at the horizon interval. The change of $\chi_i$ across the horizon interval can
be shown to be given by
\ben\label{jump}
\frac{1}{8}(2\pi)^{D-4}\ \chi_i(r=0,z) \ \bigg|_{z_h}^{z_{h+1}} = J_i \ ,
\een
where $J_i$ is the angular momentum for the rotational plane of the $i$-th rotational
Killing field $\psi_i$, expressible as
\ben
J_i = \frac{1}{8\pi} \int_\infty \star_D \D \psi_i \ .
\een
These conditions on $f_{ij}, \chi_i$, and hence $\Phi$, can be thought of as
Dirichlet type boundary conditions at $r=0$. However, these boundary conditions alone do not
guarantee that the corresponding spacetime metric~\eqref{wp} is smooth along the
axis or the horizon. For this, one needs further conditions. On each interval
$z \in (z_J, z_{J+1})$ representing an axis, and can relatively easily see that
it is required that
\ben
\lim_{r \to 0} r^{-2} \e^{\nu(r,z)} \ \sum_{i,j} f_{ij}(r,z) v_{iJ} v_{jJ} = 1 \ ,
\een
to avoid a conical singularity (``strut'') in the spacetime metric~\eqref{wp}.
On the horizon interval $z \in (z_h, z_{h+1})$, we need instead (see Appendix of~\cite{HollandsYazadjiev08b}):
\ben
\lim_{r \to 0} \e^{\nu(r,z)} \det f(r,z)^{-1} = \kappa^2 \ ,
\een
where $\kappa$ is the surface gravity of the horizon,
again, to avoid a conical type singularity. These conditions may be thought of,
coarsely speaking, as constraining the ``normal derivative'' of the fields $f_{ij}, \chi_i$,
hence $\Phi$, on the boundary $r=0$. Since one is normally not free to impose both
Dirichlet type and Neumann type boundary conditions simultaneously, this explains
intuitively the difficulty in obtaining regular solutions via the solution generating
techniques. Also note that, even if we do satisfy these conditions, they will not
imply that the metric is smooth, but only that the curvature is continuous. To get smoothness,
one would effectively have to control the full asymptotic expansion of the fields $\Phi$ near
$r=0$. Another, related, difficulty is to control the asymptotic conditions for $r \to \infty$
in order to obtain a metric $g$ with the desired asymptotic behavior near infinity.

\subsection{Divergence identities}\label{sec:divergence}

The last step towards black hole uniqueness in the stationary case are ``divergence identities''.
The idea behind such identities is actually quite simple, and is very easily explained for the toy example of
 a single scalar field, $\phi$, with strictly convex potential, $\V(\phi)$, on a bounded set
$\Omega$ in $N$-dimensional Euclidean space $\mr^N$. Such a field is described by the action
\ben
S = \int_\Omega \left( \half (\partial^i \phi) \partial_i \phi + \V(\phi) \right) \ \D^N x \ .
\een
Consider two solutions $\phi_1,\phi_2$ to the corresponding Euler Lagrange equations and define the ``current''
$j^i = (\phi_1 - \phi_2) \partial^i (\phi_1 - \phi_2)$. Taking a divergence of the current and using
the equations of motion, one has
\ben
\partial_i j^i = \partial_i (\phi_1-\phi_2) \partial^i(\phi_1 - \phi_2) + (\phi_1 - \phi_2)( \ \V'(\phi_1) - \V'(\phi_2) \ ) \ge 0 \ ,
\een
where the key $\ge 0$ relation follows from the fact that $\V'(\phi_1)-\V'(\phi_2)$ has the same sign
as $\phi_1 - \phi_2$ for a convex potential. Now let $n^i$ be the normal to $\partial \Omega$,
and assume that $\phi_1 = \phi_2$ on $\partial \Omega$ (Dirichlet conditions), or alternatively, that
$\partial_n \phi_1 = \partial_n \phi_2$ on $\partial \Omega$ (Neumann conditions). Either condition implies
that $n^i j_i = 0$ on $\partial \Omega$, so if we integrate the above equation over $\Omega$ and use
Gauss' theorem, we get the relation
\ben
0 = \int_\Omega \partial_i (\phi_1-\phi_2) \partial^i(\phi_1 - \phi_2) + (\phi_1 - \phi_2)( \ \V'(\phi_1) - \V'(\phi_2) \ ) \D^N x \ .
\een
However, the integrand is $\ge 0$, so it must be equal to zero. Thus, we conclude that $\phi_1-\phi_2$ is
a constant, and since $\V$ is {\em strictly} convex, that constant must vanish. Thus, the solution is unique with either
Dirichlet- or Neumann boundary conditions.

In our case, we do not have a single scalar field with a convex potential, but we have a field $\Phi$ without potential valued in the curved
target space $X = SL(D-2)/SO(D-2)$, satisfying the
sigma-model equations given in the last section. What replaces the {\em convexity} condition on the {\em potential}
$\mathcal V$ in this case is that the {\em metric} $\mathcal G$ on the target space $X$ has {\em negative sectional curvature}.
There are actually several ways to take advantage of this and thereby to generalize
the above divergence identity idea to this case, but they all rely, directly or indirectly, on this fact.
The first construction, the so called ``Mazur identity''~\cite{Mazur84}, is the more explicit one, and relies
on the coset structure of the target space $X$. It works as follows.

\medskip
\noindent
{\bf a) Mazur identity:} We first define an expression representing the ``difference'' between the two metrics
$g_1, g_0$ in the form~\eqref{wp},
represented by the symmetric, positive, uni-modular matrices $\Phi_1, \Phi_0$. One possibility is to choose
\ben\label{sigmadef}
\sigma := {\rm Tr}(\Phi_0^{-1} \Phi_1^{} - I) \ .
\een
Let ${\mathcal J}_i = \Phi_i^{-1} \D \Phi_i$ be the matrix currents as in the previous section, and let
$N$ be the matrix valued 1-forms defined by
\ben
N= S_1^{-1}({\mathcal J}_1^{} - {\mathcal J}_0^{})^T S_0^{} \ ,
\een
where
$S_i$ are real matrices such that $\Phi_i = S_i^{} S_i^T$. Then a calculation using the equations of
motion for $\Phi_0,\Phi_1$ gives Mazur's identity
\ben\label{mazur}
\frac{1}{r} \star_2 \D(r \star_2 \D \sigma) = {\rm Tr}(N^T \cdot N) \ge 0 \ ,
\een
which is a divergence identity of the same nature as above, having again---very importantly---a non-negative term on the right side. The domain
in question is the upper half plane $\Omega=\hat \M= \{(r,z) \mid r>0\}$, with $\star_2$ referring to the
standard flat Riemannian metric on this space. The divergence identity may now be exploited, in principle, in the same way as
for the scalar field case above, by integrating it over $\Omega$. The aim is to show that the boundary
term arising from the left side vanishes identically, which then immediately gives $N=0$ in $\Omega$.
The last fact is furthermore relatively easily seen to give $\Phi_1 = \Phi_0$, which, as is described in sec.~\ref{sec:unique}, in turn
implies that the metrics $g_1$ and $g_0$ are the same.

However, there are a number of important differences to the simple scalar field example that need to be taken into account here.
These have to do with the facts that (a) the domain $\Omega$ is not compact, and (b) the singular nature of the
boundary conditions to be considered at $\partial \Omega = \{r=0\}$. For point (a), one needs to understand
more precisely the asymptotic conditions satisfied by $\Phi_i$ for $r,z \to \infty$--here asymptotic
 expansions need to be made---and for (b)
one has to understand the behavior of $\Phi_i$ near the boundary $r=0$~\cite{HollandsYazadjiev08b}. We will discuss these issues
in the next section~\ref{sec:unique}.

Another approach, used by Bunting~\cite{Bunting83}, exploits the properties of harmonic maps in
negatively curved target spaces.

\medskip
\noindent
{\bf b) Bunting's method:} Unlike Mazur's identity, this method does not use the explicit representation of $X$ as a coset manifold, but it uses only that the metric
${\mathcal G} = -{\rm Tr}(\Phi^{-1} \D \Phi \otimes \Phi^{-1} \D \Phi)$ on $X$ is Riemannian and has negative sectional curvature. By this, one means the following. Let ${\mathcal R}_{ABCD}$ be the Riemann tensor of ${\mathcal G}_{AB}$. The Riemann tensor is
always anti-symmetric in $AB$ and $CD$, and symmetric under the exchange of $AB$ with $CD$. Thus, it can be
viewed as a symmetric, bilinear map ${Riem}: \wedge^2 T_\Phi X \times \wedge^2 T_\Phi X \to \mr$ in each tangent
tangent space of $X$. We say that $(X, {\mathcal G})$ has negative sectional curvature (or simply, is ``negatively curved'') if this
bilinear form only has negative eigenvalues. In other words, there is a $\lambda > 0$ such that, for any anti-symmetric 2-tensor $\omega$ we have $Riem(\omega, \omega) \le -\lambda \|\omega \|^2$, or in components,
\ben
{\mathcal R}_{ABCD} \omega^{AB} \omega^{CD} \le -\lambda \ {\mathcal G}_{AC} {\mathcal G}_{BD} \omega^{AB} \omega^{CD} \ .
\een
This condition replaces the condition of strict convexity for the potential in the above scalar field example. It implies, but is much more stringent than, that the scalar curvature ${\mathcal R}<0$. Let $d: X \times X \to \mr$ be the geodesic distance function on $X$, and define $d(x) = d(\Phi_1(x),\Phi_0(x))$, $x \in \hat \M = \{(r,z) \mid r> 0\}$. Then from the equations of motion~\eqref{ref}
for $\Phi_0, \Phi_1$, there follows the inequality
\ben\label{ddef}
\frac{1}{r} \star_2 \D (r \star_2 \D d) \ge 0 \ ,
\een
which is a version of the known~\cite{Jost} fact that the distance function between harmonic maps
into a negatively curved target space is sub-harmonic. The origin of this identity can be motivated as follows.
Suppose $\Phi_t: \Omega \to X$ is a family of solutions to the equations of motion~\eqref{ref}, and
let $\delta \Phi = \frac{\partial}{\partial t} \Phi_t: \Omega \to \Phi_t^* TX$ be the linearization at any fixed $t$, interpreted as the infinitesimal displacement of the 2-dimensional ``worldsheet'' in $X$ swept out by $\Phi_t$. Then, one can derive from the sigma-model field equation satisfied by each $\Phi_t \equiv \Phi$ the following equation for
$\delta \Phi_t \equiv \delta \Phi$:
\ben\label{variation}
\frac{1}{r} \star_2 \nabla (r \star_2 \nabla \delta \Phi^A) = {\mathcal R}^A{}_{BCD}(\Phi) \ (\D \Phi^B) \cdot (\D \Phi^D) \ \delta \Phi^C \ .
\een
Here $\nabla$ is the natural derivative operator in the bundle $\Phi^* TX \to \Omega$ that is inherited from
the derivative operator of the metric $\mathcal G$ on $X$.
This equation may be viewed as the generalization of the ``geodesic deviation equation'' in $X$ from curves to
surfaces, with $t \mapsto \Phi_t$ playing the role of a 1-parameter family $c_t(\lambda)$ of geodesic curves
in $X$ interpolating between two curves $c_0, c_1$, with $\delta \Phi_t$ playing the role of the displacement vector field $v_t = \delta c_t = \frac{\partial}{\partial t} c_t$ of the family, with $\D \Phi_t$ playing the role of $\D c_t/\D \lambda$, and with
$\Omega$ playing the role of the proper length parameterization of the curve $c(\lambda)$:
\ben
\frac{\D^2}{\D \lambda^2} v^A = {\mathcal R}^A{}_{BCD}(c) \ \frac{\D c^B}{\D \lambda} \frac{\D c^D}{\D \lambda}  \ v^C \ .
\een
Now assume that $[0,1] \owns t \mapsto
\Phi_t$ connects\footnote{The implicit assumption is that the space of solutions is connected. This is not
required for~\eqref{ddef}.} two given solutions $\Phi_0, \Phi_1$ of the sigma model equations. Define
\ben
s(\Phi_0, \Phi_1)(x) := \int_0^1 \left( {\mathcal G}_{AB}(\Phi_t) \delta \Phi_t^A \delta \Phi_t^B \right)(x)  \ \D t \ .
\een
Then it is straightforward to derive from eq.~\eqref{variation} the following formula
\bena
&&\frac{1}{r} \star_2 \D (r \star_2 \D  s) \\
&=& \int_0^1 \D t \left( {\mathcal G}_{AB}(\Phi_t) \ \nabla \delta \Phi_t^A \cdot \nabla \delta \Phi^B_t - {\mathcal R}_{ABCD}(\Phi_t) \ (\delta \Phi_t^A \D \Phi_t^C ) \cdot (
\delta \Phi_t^B \D \Phi_t^D)
\right) \non\\
&\ge& \int_0^1 \D t \left( {\mathcal G}_{AB} \ \nabla \delta \Phi_t^A \cdot \nabla \delta \Phi^B_t + \lambda \ {\mathcal G}_{AB} {\mathcal G}_{CD} \ (\delta \Phi_t^{[A} \D \Phi_t^{C]} ) \cdot (
\delta \Phi_t^{[B} \D \Phi_t^{D]})
\right) \ge 0 \ , \non
\eena
where to get the key $\ge 0$ relations we have used (a) that the target space is Riemannian, and (b) that it
is negatively curved. Note the close analogy between this equation, and the Mazur identity~\eqref{mazur},
and~\eqref{ddef}.
Either identity can be used
to show that the solutions $\Phi_0, \Phi_1$ are equal, by integrating over $\Omega$. Then, if one can show
that the surface term from the left side vanishes, so does $\nabla \delta \Phi_t^A$, from which one
then concludes that $\delta \Phi_t = 0$, i.e. $\Phi_t$ is independent of $t$ and hence $\Phi_0=\Phi_1$.

However, note that the method based on~\eqref{ddef} is potentially more generally applicable than Mazur's identity because it only uses that $X$ is a negatively curved manifold, whereas Mazur's identity essentially uses the fact that $X=SL(D-2)/SO(D-2)$ is a symmetric space of ``non-compact type''\footnote{
By this one means a triple $(G,H,\tau)$, where $\tau$ is an involution on the non-compact Lie-group $G$, and where
$H$ is a maximally compact subgroup invariant under $\tau$. In the present example, $\tau$
is given by $\tau(g) = (g^{-1})^T$. Mazur's identity can be generalized to any such symmetric space, as described in the original paper~\cite{Mazur84}.}. While this is a sufficient condition for $X$ to be negatively
curved by general results of symmetric spaces of non-compact type~\cite{Helgason}, it is not a necessary condition. It can be very complicated, and requires considerable creativity, to find a coset-space parameterization of this nature for the
target space $X$ (see below sec.~\eqref{sec:other}) in more general Einstein-matter systems. On the other hand, it is essentially a mechanical task to check whether or not the target space metric is negatively curved. Thus, it appears that Bunting's
method has some potential advantage of Mazur's method in this regard.

\subsection{Uniqueness theorems for rotating stationary black holes}\label{sec:unique}

We can now combine the results in the previous subsections and state a uniqueness theorem for higher dimensional
rotating, stationary black holes in vacuum general relativity. To prove this uniqueness theorem, one needs
to assume the existence of $D-3$ Killing fields $\psi_1, \dots, \psi_{D-3}$ generating an isometric action of
$U(1)^{D-3}$ on the spacetime. Note that this is very likely non-generic; the rigidity theorem is consistent
with having only a single $U(1)$ factor. One also assumes that the spacetime is asymptotically Kaluza-Klein, in the sense
described in sec.~\ref{sec1}.

\begin{thm}\label{thm4} (``Uniqueness theorem for rotating Kaluza-Klein black holes''~\cite{HollandsYazadjiev08b})
There can be at most one stationary, single horizon, vacuum, non-extremal, asymptotically Kaluza-Klein spacetime $(\M,g)$
with $s=3,4$ or $5$ large dimensions,
with $D-3$ axial Killing fields, satisfying the technical assumptions stated in sec.~\ref{sec1},
for a given interval structure $\{ \underline{v}_I, l_I \}$ and a given
set of angular momenta $\{J_i\}, i=1, \dots, D-3$.
\end{thm}
This theorem is essentially the only uniqueness theorem of this generality known for stationary
black hole solutions in higher dimensions, although the argument can presumably be generalized to
other theories of the type described below in sec.~\ref{sec:other}. For antecedents of the theorem
in 5-dimensions see~\cite{Morisawa:04} (trivial topology and horizon $S^3$) and~\cite{HollandsYazadjiev08} (arbitrary interval structure). It is not hard to generalize the theorem to black holes with multiple horizons. In that case, one has to specify separately
the angular momenta $J_i(B_j) = 1/8\pi \ \int_{B_j} \star_D \D \xi_i$ for each connected component of the horizon.

In $D=4$ with no extra dimensions, the only non-trivial interval structure for a single black hole spacetime is given by the intervals $(-\infty, -z_0],[-z_0, z_0], [z_0, \infty)$. The middle interval corresponds to the horizon, while the
half-infinite ones to the axis of the rotational Killing field. The interval vectors $\{\underline{v}_I\}$
are 1-dimensional integer vectors in this case and hence trivial. For each $z_0>0$ and for each
angular momentum $J$, there exists precisely one solution given by the appropriate member of the Kerr-family of
metrics.
Thus, one finds that the Kerr metrics exhaust all possible stationary, axially symmetric single black hole
spacetimes (satisfying the technical assumptions stated in sec.~\ref{sec1}). This is of course
just the classical uniqueness theorem for the Kerr-solution~\cite{Bunting83,Carter71,Mazur84,robinson,H72}, see~\cite{CC08,CCH12} for a coherent exposition straitening several technical issues [see also \cite{CCH12}]. The mass $m$ of the non-extremal Kerr solution characterized
by $z_0, J$ is related to these parameters by $z_0=\sqrt{m^2-J^2/m^2}>0$. Hence the uniqueness
theorem could be stated equivalently in terms of $m$ and $J$, which is more commonly done. Note that the
length of the horizon interval, $l_h = 2z_0$ tends to zero in the extremal limit, in accordance with~\eqref{lheq}.

In higher dimensions, one may similarly derive relations between the interval structure and angular
momenta on the one side, and the other invariants on the other side for any
given solution. For example, the relation between the length of the horizon interval $l_h$, the horizon
 area $A_h$, and the surface
gravity is always~\cite{HollandsYazadjiev08b}
\ben\label{lheq}
(2\pi)^{D-3} \ l_h = \kappa \ A_h \,\,\,  ,
\een
 Other formulae of this nature are provided for the Myers-Perry or black-ring solutions e.g. in~\cite{Harmark04}, but they are not expected to be universal. Of course, for most interval structures it is not known whether there actually
exists a solution, so in this sense much less is known in higher dimensions than in $D=4$.

\medskip
\noindent
{\em Outline of proof of thm.~\ref{thm4}:} We have already presented individually most of the pieces of the argument,
and we now put them together.
Suppose one has two spacetimes $(\M_1, g_1), (\M_2, g_2)$ as in the theorem. Since the interval structures
determine the topology as well as the nature of the action of the isometries by the ``orbit space theorem'', these must coincide for both
spacetimes, so $\M_1 = \M_2$. Next, by the ``Weyl-Papapetrou theorem'', one can introduce coordinates~\eqref{wp} in both spacetimes, which give
rise to sigma model fields $\Phi_1, \Phi_2$ on the upper half plane $\{(r,z) \mid r >0\}$. Next one uses the
Mazur identity~\eqref{mazur}. It is technically convenient to rewrite this equation as follows. Introduce
an auxiliary $\mr^3$ with coordinates $(x,y,z)$, related to the upper half space coordinates by
$(x,y,z)=(r \cos \varphi, r \sin \varphi, z)$. Then, view $\Phi_i$ as cylindrically symmetric functions on this
$\mr^3 \setminus \{r=0\}$. The Mazur identity implies
\ben
(\partial_x^2 + \partial_y^2 + \partial_z^2) \sigma \ge 0 \ .
\een
The key point is now to understand the behavior of $\sigma$, eq.~\eqref{sigmadef}, near the ``$z$-axis'' $\{r=0\}$
of $\mr^3$. Here, one uses again the interval structure, together with the boundary conditions~\eqref{bndy1}, and the information about the angular momenta, together with the
jump relation~\eqref{jump}. One can show from this that $\sigma$ is uniformly bounded. That part of the analysis is actually by far
the most tedious one, since $\sigma$ can have a priori very direction-dependent limits along the $z$-axis, due to the fact that the
Weyl-Papapetrou coordinates $r,z$ give a very distorted perspective on the true spacetime geometry there. Therefore these limits, and that at infinity, need to be investigated with a certain amount of care, but in the end one finds that $\sigma$ is uniformly bounded. It also follows from the definition~\eqref{sigmadef} that $\sigma \ge 0$. Now one can apply Weinstein's lemma~\cite{weinstein}, which is a version of the maximum principle:

\begin{lemma}
(``Weinstein's lemma'')
Let $\sigma(x,y,z) \ge 0$ be a continuous function on $\mr^3 \setminus \{r=0\}$ which is a solution to
$(\partial_x^2 + \partial_y^2 + \partial_z^2) \sigma \ge 0$, in the distributional sense,
and which is uniformly bounded by a constant. Then $\sigma = 0$.
\end{lemma}

Once we know that $\sigma=0$, we can derive that $\Phi_1=\Phi_2$, and
that the remaining functions $\nu,w^i$ in the Weyl-Papapetrou form also coincide for both metrics.

\subsection{Uniqueness for non-rotating (static) black holes}\label{sec:static}

While the uniqueness theorem for rotating higher dimensional black holes given in the previous
subsection required by-hand assumptions about the isometry group which are most likely
not generic, the situation is much better with regard to non-rotating stationary black holes, as we
will explain in this subsection for the case of vacuum relativity.

First, in the non-rotating case, one can apply the staticity theorem of Sudarsky and Wald~\cite{SW92},
which straightforwardly generalizes to any dimension:

\begin{thm}
(``Staticity theorem'') Let $(\M, g)$ be a smooth, asymptotically flat non-rotating, non-extremal black hole satisfying the
vacuum Einstein equations. Then $(\M, g)$ is in fact static, i.e. the Killing vector field $t$ is surface orthogonal
$t \wedge \D t = 0$.
\end{thm}

\noindent
{\bf Remark:} Versions of the staticity theorem also exist in other theories such as Einstein-Maxwell-Dilaton~\cite{Rogatkostatici}, and also to Einstein-Yang-Mills theory under certain
assumptions about the electric/magnetic charges of the fields~\cite{SW92}. \\

\medskip
\noindent
Once one knows that the spacetime is static, one can analyze its uniqueness properties by a
method that is entirely different from reduction to a sigma-model described in sec.~\ref{sec:sigma}. This method is
more powerful--albeit restricted to the static case--in that one does not have to know, a priori, that the
spacetime has any other symmetries apart from time-translations, and one can also completely bypass the topology
and rigidity arguments that are a prerequisite in the stationary case. The theorem is~\cite{Hwang98, GIS1, GIS2}:

\begin{thm}
(``Uniqueness of static black holes'')
An asymptotically flat (in the standard sense) static, vacuum black hole
with non-degenerate event horizons in any dimensions $D \geq 4$
is isometric to the Schwarzschild-Tangherlini metric.
\end{thm}
\noindent
{\bf Remark:} In $4$-dimensions, one can remove the condition that the
event horizon has no degenerate components~\cite{CRT06}.
For the higher dimensional case, the non-existence of such a static,
vacuum degenerate black holes has been discussed \cite{CRT06}
but not proven yet.

\medskip
\noindent
The theorem has been established first in $D=4$ by
Israel \cite{Israel67wq}. Later, Bunting and Masood-ul-Alam~\cite{BM},
invented an ingenious method based on a clever use of the positive mass theorem.
However, both proofs used some geometric properties that hold only in $4$-dimensions; the proof by~\cite{Israel67wq} applies
the Gauss-Bonnet theorem, while the proof of~\cite{BM} uses special
properties of the Weyl-Bach tensor in low dimensions. It is therefore not a
straightforward task to generalize the proof of either~\cite{Israel67wq} and/or~\cite{BM} to higher dimensions.
The proof by~\cite{Hwang98,Gibbons02b} is based on \cite{BM}, and it bypasses the use of the Weyl-Bach tensor using properties of special surfaces
in $\mr^n$.

\begin{center}
\begin{tikzpicture}[scale=1.1, transform shape]
\draw (0,0) -- (-2,2) -- (-4,0)  -- (-2,-2) -- (0,0);
\draw (4,0) node[right]{$i_0 \cong S^{D-2}$};
\draw (-4,0) node[left]{$\overline{i_0} \cong S^{D-2}$};
\draw (0,0) -- (2,2) -- node[above right] (c) {$\I^+$}  (4,0) -- node[right] (a) {$\I^-$} (2,-2) -- (0,0);
\shade[left color=gray] (0,0) -- (-2,2) decorate[decoration=snake] {-- (2,2)} --  (0,0);
\draw (0,0) -- (-2,2) decorate[decoration=snake] {-- (2,2)} -- (0,0);
\shade[left color=gray] (0,0) -- (-2,-2) decorate[decoration=snake] {-- (2,-2)} -- (0,0);
\draw (0,0) -- (-2,-2) decorate[decoration=snake] {-- (2,-2)} -- (0,0);
\draw[->, very thick] (2,3.2) node[above right]{singularity} -- (1, 2.2);
\draw[very thick, black] (0,0) -- (4,0);
\draw (2,0) node[below]{$\Sigma$};
\draw[very thick, black] (0,0) -- (-4,0);
\draw (-2,0) node[below]{$\overline{\Sigma}$};
\filldraw (0,0) circle (.05cm);
\filldraw (4,0) circle (.05cm);
\draw[->, very thick] (-2,3) node[above left] {BH $=\M \setminus J^{-}(\mathscr{I}^{+})$} -- (-.5,1.5);
\draw (1.3,1.3) node[black,above,left]{$\H^+$};
\draw (1,-1) node[black,below,left]{$\H^-$};
\draw(0,0) node[black, below]{$B$};
\end{tikzpicture}
\end{center}

\noindent
That proof 
may be summarized as follows (see also \cite{IIS11}):
\begin{enumerate}
\item First, one constructs a complete Riemannian manifold
$\tilde \Sigma$ with an asymptotically flat metric $\tilde h$.
This is done by taking a spatial slice $(\Sigma,h)$ of the spacetime,
eq.~(\ref{metric:static}), orthogonal to the orbits of
the Killing vector field, $t = {\partial}/{\partial \tau}$, with
$\partial \Sigma=B$, and then doubling $(\Sigma,h)$ across
the horizon cross section $B$ and gluing a copy $\overline \Sigma$ onto $\Sigma$ along $B$.
Then, one performs a conformal rescaling from $h$ to $\tilde h = \Omega^2 h$
on the doubled spacetime $\tilde \Sigma = \Sigma \cup \overline \Sigma$,
with $\Omega$ chosen in such a way that $\Omega \to 1$ near the
spatial infinity $i_0$ of $\Sigma$, and $\Omega \to 0$ near the spatial
infinity $\overline i_0$ of $\overline \Sigma$, and adds a single point
to $\overline \Sigma$ to cap off $\overline i_0$. Hence, the
slice $\tilde \Sigma$ only has one asymptotic end, $i_0$, see the above figure.

\item
The next step is to show that $(\tilde \Sigma, \tilde h)$ is flat Euclidean space.
This is done by an appropriate choice of the conformal factor $\Omega$,
the asymptotic flatness of the original metric $h$, and the use
of the positive energy theorem.
To make this work, $\Omega$ must be chosen to make
the scalar curvature $\tilde R$ of $\tilde h$ non-negative.
To start, note that the condition of the asymptotic flatness implies:
\bena
   N &=& 1 - \frac{M }{r^{D-3}} + O(r^{-(D-2)})  \ , \\
   h &=& \left( 1 + \frac{2}{D-3}\frac{ M }{r^{D-3}} \right) \delta
      + O(r^{-(D-2)})
\label{condi:AF}
\eena
as in eq.~(\ref{metric:static}).
Now, choose the conformal factor $\Omega=\{ \Omega_+ \mbox{on $\Sigma$}, \,
\Omega_- \mbox{on $\bar{\Sigma}$} \}$ by
\ben
  \Omega_\pm = \left(\frac{1 \pm N}{2}\right)^{2/(D-3)} \,.
\label{def:Omega}
\een
Then, asymptotic flatness, eq.~(\ref{condi:AF}), translates
in terms of the rescaled metric $\tilde h$ into
\ben
 \tilde h = \delta + O(r^{-(D-2)}) \, ,
\een
implying, very importantly, that the mass $\tilde M$ of $(\tilde \Sigma, \tilde h)$ is zero.
One can also check that with the choice eq.~(\ref{def:Omega}),
the scalar curvature of $\tilde h $ on $\tilde \Sigma$ is non-negative, and
that it is sufficiently regular across $B$ in order to be able to apply the
positive mass theorem~\cite{PET, PET2, hawkinghorowitz}, which states that:

\begin{thm}
Consider an $n$-dimensional, complete, spin\footnote{The requirement that $\tilde \Sigma$ be spin is not needed in dimensions $n=3$
because the technique of~\cite{PET} does not require spinors. A generalization
of that argument to dimensions less than $n=8$ is also available, so up to that
dimensions, we do not need a spin structure. In higher dimensions, the proof of the
theorem is based on the Witten spinor method, and therefore only works for spin manifolds.}, Riemannian manifold,
$(\tilde \Sigma, \tilde h)$ with asymptotically flat end,
i.e., Euclidean outside a compact set and the metric decays as
in eq.~(\ref{condi:AF}).
If the scalar curvature of $\tilde h$ is non-negative, then the ADM mass
of $(\tilde \Sigma,\tilde h)$ is also non-negative. Furthermore,
if the ADM mass vanishes, then $(\tilde \Sigma, \tilde h)$ is isomorphic
to flat space $(\mr^n, \delta)$.
\end{thm}

By the positive mass theorem, we conclude that $\tilde h$ is flat, and hence that
$h$ is conformally flat, i.e. $h=\Omega^{-2} \delta$, and we also conclude that
$\tilde \Sigma \cong \mr^{D-1}$, i.e., $\Sigma \cong \mr^{D-1} \setminus {\rm BH}$, where ${\rm BH}$
is a compact manifold with boundary $B$. We also know by construction that
the lapse function $N$ is determined by the conformal factor $\Omega$
via eq.~(\ref{def:Omega}). Thus, $(\Sigma,h,N)$ will be unique if we can show that
$\Omega$ is uniquely specified.

\item One shows that $\Omega$ is unique.
Introducing $u_\pm := \Omega_\pm^{-1}$, one shows that
the Laplace equation holds on the flat base space $\mr^{D-1}$,
\ben
   \Delta u_\pm = 0 \,,
\label{eq:Laplace}
\een
where $\Delta$ is the ordinary flat space Laplacian for the metric $\delta$.
The boundary condition at infinity $r \rightarrow 0$
follows from the conditions of asymptotic flatness on $(\Sigma,N,h)$, eq.~(\ref{condi:AF}).
The boundary condition at $B$ follows from the fact that $B$ represents a horizon, eq.~(\ref{def:Omega}),
which gives $N=0$ on $B$. These conditions together with the Laplace equation
uniquely determine $u_\pm$, hence $\Omega$, once
we know exactly what the nature of the boundary $B$ in our flat base space $\mr^{D-1}$ is.

\item The final step is to show that the inner boundary $B$
is a spherically symmetric hypersurface in the base space $\mr^{D-1}$. It is then straightforward to show
that the unique solutions $h$ and $N$ coming from eq.~(\ref{eq:Laplace}),
which also must respect the spherical symmetry,
in fact correspond to the lapse function and the spatial part of
the Schwarzschild metric in isotropic coordinates.
At this point, it becomes relevant what the spacetime dimension is.
In the $4$-dimensional case, spherical symmetry can be shown
using the fact that Weyl-tensor of $3$-dimensional space
vanishes. This argument does not work in the case in $D-1 \geq 4$.
However, in any dimension, the vacuum Einstein equation implies that
the lapse function $N$ is harmonic on $(\Sigma,h)$, from which one
can argue, using the maximum principle, that it is possible to take $N$ as a coordinate, so that
\ben
  h = \rho^2 \D N^2 + \gamma_{ab}\D x^a \D x^b
\een
where the function $\rho$ has been introduced so that the one-form
$\rho \D N$, normal to each $N=const.$ level surface with the metric
$\gamma_{ab}$, is normalized with respect to the metric $h$, and
the trajectory of the coordinates $x^a$ on level sets of $N$ are
orthogonal to each level set. The level set $N=0$ corresponds to the bifurcation surface $B$ of the horizon.
The regularity of $g$ and the bifurcate surface property implies that
$B$ must be totally geodesic in $\Sigma$, i.e.,
the extrinsic curvature $k$ of $B$ as an embedded $(D-2)$-hypersurface
in $\Sigma$ must be zero. Using this result, one can show that
the conformally transformed $B$ viewed as a surface in $(\tilde \Sigma,\delta)$ must be
totally umbillic, i.e., ${\tilde k}_{ab} \propto {\tilde \gamma}_{ab}$.
Then, one can appeal to a well-known mathematical result that a totally
umbilical embedding of $(n-1)$-surface into $n$-dimensional Euclidean space
is spherical \cite{KN5-1}. Thus, each connected component of
the horizon $B$ has been shown to be spherically symmetric.
One can also show that $B$ must be connected, therefore $B$ is
metrically a sphere in $(\mr^{D-1}, \delta)$. As we have already
seen, this implies that $(N,h)$ coincide with the corresponding
quantities in the Schwarzschild spacetime.

\end{enumerate}

\subsection{Extremal black holes and their near horizon geometries}

The uniqueness theorems discussed so far make the assumption that
all components of the event horizon be non-degenerate. There are, however,
a number of known exact solutions with degenerate (extremal) event horizon. It is interesting to know
to what extent these solutions are unique.

\begin{enumerate}

\item[a)]
\underline{Rotating case :} The uniqueness proof for
rotating black holes proceeds, as in the non-degenerate case, by reducing the problem to a boundary
value problem for harmonic maps, using either Mazur's or Bunting's method.
As in the non-degenerate case, the last step is to analyze boundary conditions for those
harmonic maps on the two-dimensional orbit space $\hat \M = \{(r,z) \mid r>0\}$.
One of the boundary conditions is imposed at the
horizon interval on the boundary $r=0$.
However, a new feature arises when the horizon is the degenerate type, because
the corresponding interval shrinks to a single point according to~\eqref{lheq}. This means
basically that the Weyl-Papapetrou coordinate system is very ill-adapted to
resolve the geometry near the horizon, and some sort of ``blow up'' is required
at this point in order to analyze the boundary behavior for the harmonic maps there.

In 4-dimensions, uniqueness theorems for extremal Kerr and extremal
charged Kerr black holes have been shown
recently by \cite{CN10,AHMR10,FL09}. The key new element in these proofs is
a uniqueness result~\cite{Hajicek74,LP03,KL08} that the
near-horizon geometry for a degenerate Killing horizon of
any stationary axisymmetric vacuum spacetimes with given mass
and angular momentum must agree with
that of the extremal Kerr metric~\cite{BardeenHorowitz99}.
The near-horizon geometry is obtained by an infinite scaling of the horizon
neighborhood, which, loosely speaking, achieves the desired blow up at the
horizon, and which makes it possible to understand the
relevant boundary conditions for the sigma-model fields at the degenerate horizon.
(A more detailed description of this scaling is given below.)
A similar uniqueness result also holds for the extremal electrovacuum black
hole case.
It is possible also in $D \ge 5$ dimensions to
classify the near horizon geometries under the type of symmetry assumption as in
our uniqueness theorem~\ref{thm4}, see below.
However, these results alone do not suffice for a proof of
an analogue of our uniqueness theorem~\ref{thm4} in general dimension,
because it is unclear in general dimension how the parameters of the weighted orbit
space, $\{l_J\}$ and $\{ \underline{v}_J \}$, are related to the
parameters of the near horizon geometries. A result in this
direction in $D=5$ is Theorem~2 of Ref.~\cite{FL09}, stating
that the interval structure can be used to uniquely determine extremal vacuum
black holes in $5$-dimensions, (as well as near-horizon geometries)
under similar assumptions of the uniqueness theorem of~\cite{HollandsYazadjiev08}.
In general dimensions, however, this remains an open problem.

\item[b)]
\underline{Static case :}
As shown in \cite{C99,CRT06}, there are no asymptotically flat,
static vacuum black holes with degenerate components of the event horizon.
This analysis has been generalized to include Maxwell field \cite{CT07}
by using properties of the near-horizon geometry. It was shown that
the Majumdar-Papapetrou spacetime is the only asymptotically flat, static
electro-vacuum black hole spacetime with degenerate components
of the event horizon \cite{CRT06}.
See also Ref.~\cite{Rogatko03,Rogatko05} for related work.
Static near-horizon geometries in 5-dimensional Einstein-Maxwell
theory with Chern-Simons term have been classified \cite{KL09b}.

\item[c)]
\underline{Supersymmetric black holes} are extremal and their classification is discussed below in sec.~\ref{sec:other}.
\end{enumerate}

As we have explained in (a), the concept of near horizon geometry has a potential
significance in proving uniqueness of higher dimensional black holes in a situation
wherein one has the amount of symmetry described in our uniqueness theorem~\ref{thm4}.
However, as we have also emphasized several times already, these symmetry assumptions are
most likely non-generic (other than in $D=4$). Therefore, as a stepping stone towards a
classification of extremal black holes having less symmetry, one might at first look at the
corresponding classification of their near horizon geometries, which are simpler, i.e., more symmetric. Let us therefore describe what a near horizon geometry actually is, what the corresponding equations are, etc.

\vspace{1cm}
\noindent
\underline{Near Horizon Geometries:} Let $(\M, g)$ be an extremal vacuum black hole
spacetime with Killing horizon $\H$, and Killing vector field $K$ tangent to the null-generators of $\H$. (In the non-extremal case, the existence of $K$ automatically follows from the stationarity assumption by the rigidity theorem, in the extremal case, the rigidity theorem is not quite as powerful as yet, see the discussion in sec.~\ref{sec:rigidity}.)
Now write the metric near $\H$ in Gaussian null-coordinates as in eq.~\eqref{gaussian}.
As argued there, we may assume that, in Gaussian null coordinates, $K = \partial/\partial u=n$. Then, none of the tensor fields $\alpha, \beta, \gamma$ in~\eqref{gaussian} depend on $u$,
but only on $r$ and the coordinates $x^a$ on the horizon cross section $B$.
 From eq.~\eqref{nnaln}, and the definition of the surface gravity, eq.~\eqref{surfgr}, it
follows that $\alpha=0$ on $\H$ in the extremal case. Hence, $\alpha$ can be written
as $r$ times a smooth function, which by abuse of notation we shall call $\alpha$ again.
In summary, in the extremal case, the metric in an open neighborhood of a Killing horizon $(r=0)$ can be assumed
to take the form
\ben\label{gaussian1}
g = 2 \ \D u( \D r - r^2 \alpha \ \D u - r\beta_a \ \D x^a) + \gamma_{ab} \ \D x^a \D x^b \ .
\een
None of the tensor fields depend on the coordinate $u$. Consider now for small $\epsilon > 0$ the diffeomorphism
$\phi_\epsilon$ mapping a point with coordinates $(r,u,x^a)$ to the point with
rescaled coordinates $(\epsilon r, u/\epsilon, x^a)$. The near horizon limit $g_0$ is defined
by the limit, as $\epsilon \to 0$, of the family of metrics $g_\epsilon = \phi_\epsilon^* g$.
It is concretely given by the {\em same} formula as~\eqref{gaussian1}, but
with the tensor fields $\alpha, \beta, \gamma$ {\em replaced by their restriction to the horizon cross section} $B$, i.e. $r=0$. In other words, these tensor fields now no longer depend on $r$
(and not on $u$ either). By construction, the near horizon limit is a vacuum solution. The infinite rescaling means that Einstein's equations reduce to the lowest order in $r$ contribution to eqs.~\eqref{gee} (and the remaining components). Concretely, they are in the vacuum
\ben\label{nhe}
\begin{split}
0 &= R_{ab}(\gamma)
         - \half \ \pounds_\beta \gamma_{ab}
         - \half \ \beta_a\beta_b \ , \\
0 &= 2 \ \alpha + \half \ \beta_a \beta^a + \half \ D_a \beta^a \ .
\end{split}
\een
Thus, the near horizon geometry limit has drastically reduced the complexity of the field equations: we `just' have to solve the above two equations for the tensor fields
$\alpha, \beta, \gamma$ on the compact $(D-2)$-dimensional horizon cross section manifold $B$. Furthermore, $\alpha$ is absent from the first equation, which we may therefore solve first, and then trivially use the second equation to determine $\alpha$.

Unfortunately, it is still by no means a simple task to solve the near horizon equations, and
a complete classification of solutions for an arbitrary compact manifold $B$ (or at least a class of such manifolds) is not available at present. Besides the known solutions that can be obtained from concrete extremal black holes such as Myers-Perry, an infinite class of solutions was recently obtained by~\cite{KL10} for even dimensions $D$. These authors use an ansatz for $\gamma, \beta$ in terms of a $(D-4)$-dimensional compact K\" ahler-Einstein base space (inspired by the concrete form of the near horizon Myers-Perry solution), which is the basic input into their metrics. The near-horizon equations can then be solved by integrating certain ordinary differential equations. Since a wide variety of compact, even-dimensional K\" ahler-Einstein manifolds is known, this produces a correspondingly wide class of solutions to the near horizon equations~\eqref{nhe}, some of which have a very low amount of symmetry
(the authors also admit a cosmological constant, suppressed in~\eqref{nhe}). Consequently, for most of their solutions, it is not known whether they actually arise from a full-fledged
extremal black hole, or not.

\medskip

Another approach to the classification of near horizon geometries is to assume, from the outset, a comparable amount of symmetry as in our uniqueness theorem for non-extremal black holes, thm.~\ref{thm4}. Thus, we assume by hand the existence of $D-3$ additional Killing fields $\psi_1, \dots, \psi_{D-3}$ generating an isometric action of $U(1)^{D-3}$ on spacetime. We may assume that these Killing fields are tangent to the cross section manifold $B$, so $\alpha, \beta, \gamma$ are Lie-derived by these Killing fields. What is the total isometry group of such a near horizon metric? From the outset, we assumed the Killing field $K=\partial/\partial u$, but, after the near horizon limit, we obtain the additional Killing field $X=-u \ \partial/\partial u + r \ \partial/\partial r$ which generates the rescalings $\phi_\epsilon$. Together, $K,X$ generate a 2-dimensional Lie-group $G_2$ which is isomorphic to the semi-direct product of translations and dilatations of the real line. So, the total, manifest, isometry group of~\eqref{gaussian1} with our assumptions is $G_2 \times U(1)^{D-3}$.

It is surprising that Einstein's equations imply that the full isometry group is actually even bigger and in fact enhanced to $SL(2) \times U(1)^{D-3}$, where $SL(2)$ contains $G_2$~\cite{KLR07}. The origin of this symmetry enhancement can be explained as follows.
Letting $f_{ij} = g(\psi_i, \psi_j)$ as above, a coordinate $-1 \le x \le 1$ is introduced by
demanding the metric $\gamma$ can be written as
\ben
\gamma = \frac{1}{C^2 \det f} \, \D x^2 + f_{ij}(x) \D \varphi^i \D \varphi^j \ ,
\een
where $C>0$ is a constant. Geometrically, $x$ is a coordinate parameterizing the
orbit space $B/U(1)^{D-3} \cong [-1,1]$. The $2\pi$-periodic coordinates $\varphi^i$
have also been introduced in such a way that $\psi_i = \partial/\partial \varphi^i$.
It is next shown that there is a function $\lambda$ on $B$ (effectively of $x$) such that
\ben
\beta = \D \lambda + C \e^\lambda \, k_i \D\varphi^i \, ,
\een
where we have introduced the scalar functions
\ben
k_j := C^{-1} \e^{-\lambda} \, i_{\psi_j} \beta \, .
\een
The next coordinate, $\rho$, is defined by
\ben
\rho:= r \e^\lambda \, ,
\een
and we keep $u$ as the last remaining coordinate. The coordinates
$(\varphi^i, \rho, x, u)$ are the desired geometrical coordinates.
The Einstein equations now impose relations between the various functions $k_j, \alpha, \lambda, f_{ij}$, ~\cite{KLR07}. Namely, one finds that $k^i=f^{ij} k_j$ are simply {\em constants},
and that $2\alpha \e^{-\lambda} - \e^\lambda k_i k^i$
is a negative\footnote{Here one must use that the metric is {\em not} static, i.e. that not all
 $k^i$ vanish.} {\em constant}, which one may choose to be $-C^2$ after a suitable rescaling of the
coordinates $\rho,u$ and the constants $k^i$, and by adding a constant to $\lambda$. The Einstein
equations are seen to further imply that 
$
\e^{-\lambda} = (1-x^2)(\det f)^{-1} 
$.
These conditions then imply together that the near horizon metric
is given by
\ben\label{nhg}
g_0 = \frac{1-x^2}{\det f(x)}(2\D u \D \rho - C^2 \rho^2 \D u^2) + \frac{\D x^2}{C^2 \det f(x)}  +
f_{ij}(x)( \D \varphi^i + \rho Ck^i \, \D u)(\D \varphi^j + \rho Ck^j \, \D u)
\een
where $k^i, C$ are constants. The $SL(2)$ symmetry
now arises because the first term in parenthesis is just the metric of $AdS_2$, with
isometry group $SL(2)$~\cite{KLR07}. The $AdS_2$-factor plays a crucial role in
string theory analyses of black hole entropy of extremal black holes.

The functions $f_{ij}(x)$ and their relationship to the constants $k^i, C$ still have not been
determined, but it turns out that this can also be done, making use of the sigma-model formulation explained in sec.~\ref{sec:sigma} (the coordinates $x,\rho$ turn out to be
closely related to the Weyl-Papapetrou coordinates $r,z$ in thm.~\ref{wpthm}). This final step was carried out in $D$-dimensions in~\cite{HI10}, using earlier work of refs.~\cite{KL08,KL09b,FKLR08}.

To state the result, we must remember that, by the orbit space theorem~\ref{thm:orbit},
at each point $x = \pm 1$, there exists an integer linear combination of the vectors
$\psi_i$ which vanishes. The coefficients in these two integer linear combinations make up
two vectors $\underline{a}_+:=\underline{v}_{h-1}, \underline{a}_-:=\underline{v}_{h+1} \in \mz^{D-3}$, which are the ``winding numbers'' of the intervals adjacent to the horizon interval in the orbit space theorem. The classification theorem is then~\cite{HI10}:

\begin{thm}\label{thm5}
All smooth, non-static, near horizon metrics with $D-3$ commuting Killing fields
generating $U(1)^{D-3}$
are parametrized by real parameters $c_\pm, \mu_i, s_{Ii}$,
and the integers $a_\pm^i$ where $I=0,\dots,D-5$ and $i=1, \dots, D-3$,
and ${\rm g.c.d.}(a_\pm^i) = 1$.
The explicit form of the near horizon metric in terms of these parameters is
\bena\label{NH}
g_0 &=& \e^{-\lambda} (2\D u \D \rho - C^2 \rho^2 \D u^2 + C^{-2} \, \D \theta^2) +
\e^{+\lambda} \Bigg\{ (c_+-c_-)^2 (\sin^2 \theta) \, \Omega^2 \non\\
&& +(1+\cos \theta)^2 c_+^2 \sum_I \left( \omega_I - \frac{s_I \cdot a_+}{\mu \cdot a_+} \Omega \right)^2
+(1-\cos \theta)^2 c_-^2 \sum_I \left( \omega_I - \frac{s_I \cdot a_-}{\mu \cdot a_-} \Omega \right)^2\non\\
&& +  \frac{c_\pm^2\, \sin^2 \theta}{(\mu \cdot a_\pm)^2} \sum_{I < J} \Big(
(s_I \cdot a_\pm) \omega_J - (s_J \cdot a_\pm)\omega_I \Big)^2
\Bigg\} \, .
\eena
Here, $x=\cos \theta$, the sums run over $I,J$ from $0, \dots, D-5$, the function
$\lambda(\theta)$ is given by
\ben
\exp[-\lambda(\theta)] = c_+^2(1+\cos \theta)^2 + c_-^2(1-\cos \theta)^2 + \frac{c_\pm^2 \sin^2 \theta}{(\mu\cdot a_\pm)^2} \sum_I (s_{I}\cdot a_{\pm})^2 \, ,
\een
$C$ is given by $C = 4c^2_\pm[(c_+-c_-)(\mu \cdot a_\pm)]^{-1}$, and we have defined  the 1-forms
\bena
\Omega(\rho) &=& \mu \cdot \D \varphi + 4C\rho\frac{c_+c_-}{c_+ -c_-} \D u \\
\omega_I(\rho) &=& s_{I} \cdot \D \varphi +
\frac{\rho}{2} \, C^2 (s_{I} \cdot a_+ + s_I \cdot a_-)  \, \D u \, .
\eena
We are also using the shorthand notations such as $s_{Ii} a^i_+ = s_I \cdot a_+$, or
$\mu \cdot \D \varphi = \mu_i \D \varphi^i$, etc.
The parameters are subject to the constraints $\mu \cdot a_\pm \neq 0$ and
\ben\label{constraints1}
\frac{c_+^2}{\mu \cdot a_+} = \frac{c_-^2}{\mu \cdot a_-}
\, ,
\quad
\frac{c_+ (s_{I} \cdot a_+)}{\mu \cdot  a_+} = \frac{c_- (s_{I} \cdot a_-)}{ \mu \cdot  a_-}
\, ,
\quad
\pm 1 = (c_+ - c_-) \, \epsilon^{ijk \dots m} s_{0i} s_{1j} s_{2k} \cdots \mu_m
\een
but they are otherwise free. The coordinates $\varphi^i$ are $2\pi$-periodic,
$0 \le \theta \le \pi$, and $u,\rho$ are arbitrary. When writing ``$\pm$'', we mean that the
formulae hold for both signs.
\end{thm}

{\bf Remarks} 1) The number of free real parameters, after taking into account the constraints~\eqref{constraints1}, is $(D-3)(D-4)$. It is not clear that all near horizon geometries in the above theorem can be matched to known black hole solutions.\\
2) The possible horizon topologies for $B$ are encoded in the inter vectors $\underline{a}_+:=\underline{v}_{h-1}, \underline{a}_-:=\underline{v}_{h+1} \in \mz^{D-3}$, see figure~\ref{table1}. \\
3) A similar analysis for minimal supergravity in $D=5$ dimensions has been carried out, following the method of~\cite{HI10}, in~\cite{KL11}. It would be interesting to carry out
a corresponding analysis in 11-dimensional supergravity. \\
4) A similar result is not available at present for a non-zero cosmological constant, where the type of technique used in the proof does not seem to work.

\section{Other Theories in $D \ge 4$} \label{sec:other}

Our discussion of uniqueness theorems in both the rotating (stationary) and non-rotating (static) case
has been restricted to the vacuum field equations, and to specific boundary conditions at infinity (and also to non-extremal black holes). Also, our discussion of some related other results, such as the
sigma-model reduction, and Weyl-Papapetrou form, has been restricted to the vacuum theory.
It is clearly of interest to (a) consider other theories, i.e., matter fields and (b) to discuss more general asymptotic boundary
conditions. In this section, we give examples of what has been achieved in this direction, or what we think probably could be achieved without too much difficulty. The following list gives an indication of the situation.

\vspace{1cm}
\begin{center}
\begin{tabular}{c|c|c}
Theorem & Theory & Boundary Conditions \\
\hline
rigidity thm. & \eqref{genth} + null energy cond. & asympt. flat, (A)dS \\
topology thm. & dominant energy cond. & any\\
top. censorship & null conv. cond. & asympt. flat,KK,AdS(?)\\
staticity thm. & Einstein-Maxwell & asympt. flat  \\
uniqueness stationary & vacuum + $U(1)^{D-3}$ symmetry & asympt. KK  \\
& $\exists \sigma$-model formulation + $U(1)^{D-3}$ symmetry & asympt. KK  \\
uniqueness static & Einstein-Maxwell-Dilaton & asympt. flat  \\
& Einstein+sigma model & asympt. flat \\
& Einstein+form fields & asympt. flat  \\
\end{tabular}
\end{center}

As is apparent from the list, general structural theorems such as the rigidity-, toplogy-, topological censorship
theorems can be proved under fairly minimal assumptions, i.e. appropriate energy conditions, that are
verified in a wide variety of matter models (however, extremal black holes often
pose problems). On the other hand, the uniqueness theorems require more special input.
The two main gaps in our understanding here are: 1) Stationary
black holes in theories that do not admit a sigma-model reduction of the kind described above, either
because (a) lack of symmetry or (b) due to the nature of the Lagrangian. This is a very significant
shortcoming, because investigations based on approximations~\cite{Dias,blackfold} indicate that the required $U(1)^{D-3}$-symmetry is not generic, and because it excludes many theories of interest. 2) For static black holes, the uniqueness proofs work in
much greater generality, but one does not know as yet e.g. how to treat asymptotic KK-boundary conditions, and
again, extremal black holes are sometimes more difficult to deal with.
In the following two subsections, we discuss some of these points in a bit more detail.

\subsection{Rotating stationary black holes}

Let us first indicate in which kinds of higher dimensional gravity theories, other than
vacuum Einstein theory, and under what type of symmetry assumption,
one might apply the uniqueness arguments presented in sec.~\ref{sec:unique} for rotating black holes.
The bottleneck of the argument are the Weyl-Papapetrou form, orbit space theorem, and the existence of a
divergence identity of the type of the Mazur- or Bunting identity, see sec.~\ref{sec:divergence}. The Weyl-Papapetrou
form requires $D-2$ commuting Killing fields, together with the vanishing of the right side of eq.~\eqref{killid}.
This depends in general on the form of the matter stress-tensor, but is OK in general theories of the form~\eqref{genth}, plus possibly Chern-Simons type terms. However, already this step does not work e.g. for
non-abelian gauge fields. Both the Mazur- and Bunting identity are based on a sigma-model formulation of the
dimensionally reduced theory. This also requires $D-2$ commuting Killing fields, but puts much more stringent
further constraints. Examples of known  sigma-model formulations for the dimensionally reduced theory to 3-dimensions (assuming $U(1)^{D-3}$ symmetry) are given in
the following table:

\begin{center}
\begin{tabular}{c|c|c|c}
Theory & $D$ & Sigma model? & Coset \\
\hline
Einstein-Maxwell & 4 & yes & $SU(2,1)/S(U(1) \times U(2))$ \ \cite{neugebauer}\\
Einstein-Maxwell & $\ge 5$ & no & ---\\
Einstein-Yang-Mills & $\ge 4$ & no & ---\\
Vacuum & $\ge 4$ & yes & $SL(D-2)/SO(D-2)$ \ \cite{Ehlers,Maison79}\\
Vacuum $+\Lambda$ & $\ge 4$ & no & ---\\
Minimal Sugra & 5 & yes & $G_{2(+2)}/SO(4)$ \ \cite{mizo1,clement}\\
Sugra & 11 & yes & $E_{8(+8)}/SO(16)$ \ \cite{julia,mizo2}\\
Sugra & 10 & yes & $SO(8,8)/(SO(8) \times SO(8)) \ \cite{julia}$
\end{tabular}
\end{center}

Further models leading to sigma-models upon dimensional reduction are discussed in~\cite{breitgibbons}.
Note that most theories in the list which have a sigma-model formulation are related
to supergravity theories.
The coset formulation of the reduced Einstein-Maxwell system in $D=4$ is the basis of the
classic uniqueness theorem for the charged Kerr black hole already described at the beginning.
For Einstein-Maxwell theory in $D \ge 5$ dimensions, there is still a sigma-model formulation
if one makes further, by-hand, restrictions on the form of the Maxwell field, and metric, see below.

\medskip
\noindent
{\bf $D=11$ Supergravity:}
Maybe
to the most interesting case is the case of $D=11$ supergravity. This is based on the exceptional Lie-group\footnote{The subscript $(+8)$ indicates a special real form.} $E_{8(+8)}$ which has a---well-deserved---reputation
for being complicated. A parameterization of the coset was given e.g. by~\cite{mizo2} (for original papers, see~\cite{julia,breitenl}); we outline how to get a uniqueness theorem parallel to that given in thm.~\ref{thm4}. The details are in~\cite{hollands12}.

First, the bosonic part of the supergravity Lagrangian in $11$ dimensions is
\ben
S= \int_\M \half R \star_{11} 1 -  F \wedge \star_{11} F - \frac{2}{3} \ B \wedge F \wedge F \ ,
\een
where $B$ is a 3-form potential with field strength $F = \D B$. In parallel with the vacuum case,
we assume 8 commuting axial Killing fields $\psi_1, \dots, \psi_8$ in addition
to the timelike Killing field $t$, i.e. an isometric action of $U(1)^8 \times \mr$ on the spacetime. Additionally,
it is required that these vector fields Lie-derive also the potential $B$. Then it can be seen that
the orbit space theorem still applies. Furthermore, the metric takes again the Weyl-Papapetrou form, because the
right side of~\eqref{killid} vanishes. In particular, we can define the interval structure of the spacetime,
with $8$-dimensional vectors $\underline{v}_J \in \mz^8$. We can also define electric type charges by
\ben
Q_{\rm E}[C] := \int_C (\star_{11} F +  B \wedge F) \ ,
\een
for any 7-cycle $C \in H_7(\M, \mz)$. These provide further invariants that are not present in
pure gravity. To define the coset, one first has to introduce the potentials analogous to~\eqref{chieq}.
As in the vacuum case, we write the 11-dimensional metric $g$ as in~\eqref{kkansatz}, with 1-forms $A^i$, and with
$f_{ij} = g(\psi_i, \psi_j)$. Then, writing $B_i = i_{\psi_i} B, B_{ij} = i_{\psi_i} i_{\psi_j} B$
etc., one shows that scalar potentials $\varphi^{ij}, \chi_i$ can be defined by the following equations:
\bena
&&F^{ij} - \D A^k B^{ij}{}_k = - \det f^{-1} \ \star_3 \left(
\D \varphi^{ij} - \frac{1}{36} \D B_{klm} B_{npq} \epsilon^{ijklmnpq}
\right) \\
&&\D A_k = - \det f^{-1} \ \star_3 \left( \D \chi_k - B_{kij} \D \varphi^{ij} + \varphi^{ij} \D B_{kij}
+ \frac{1}{54} \epsilon^{ijlmnpqr} B_{ijk} \D B_{lmn} B_{pqr} \right) \ . \non
\eena
The effective 3-dimensional sigma-model Lagrangian for the model is then given by the same formula \eqref{Sreduced} as in pure gravity, but with a considerably more elaborate definition of $\Phi$ taking values in the coset $E_{8(+8)}/SO(16)$, which is defined in terms of the scalars $f_{ij}, \chi_i, B_{ijk}, \varphi^{ij}$. The precise definition is rather involved and was developed in~\cite{mizo2}:
The matrix $\Phi$ acts on 3-tuples of tensors of the form\footnote{This is the adjoint representation ${\bf 248}$
of $E_{8(+8)}$, which under $SL(9)$ decomposes as
${\bf 80} \oplus {\bf 84} \oplus \overline{\bf 84}$.} $(z^{M}{}_{L}, x_{[LMN]}, y^{[LMN]})$, where capital Roman
letters run from $1, \dots, 9$, where $z^I{}_I =0$, and where a square bracket denotes anti-symmetrization.  Define
\ben
V_I{}^J = \left(
\begin{matrix}
e_i{}^a & - (\det f)^\half \chi_i\\
0 & (\det f)^{\half}
\end{matrix}
\right)
\een
with $e_i{}^a e_k{}^b \delta_{ab} = f_{ik}$, and define
\bena
v_{IJK} &=&
\begin{cases}
-2\sqrt{3} B_{ijk} & \text{if $i=I, j=J, k=K$ all $\le 8$}\\
0 & \text{otherwise.}
\end{cases}
\\
w^{IJK} &=&
\begin{cases}
-6\sqrt{3} \varphi^{ij} & \text{if $i=I, j=J$ all $\le 8$ and $K=9$,}\\
0 & \text{otherwise} \ .
\end{cases}
\eena
Next, define
\bena
&&\hspace{5cm} Z = \\
&&\left(
\begin{matrix}
0 &
\tfrac{1}{6}(w^{J[MN} \delta^{L]}_I - \tfrac{1}{9} w^{LMN} \delta^J_I) &
-\tfrac{1}{6}(v_{I[MN} \delta^J_{L]} - \tfrac{1}{9} v_{LMN} \delta^J_I)\\
-3(v_{M[JK} \delta_{I]}^L - \tfrac{1}{9} v_{IJK} \delta^L_M) &
0 &
\tfrac{1}{36\sqrt{3}} \epsilon_{ILKPQRLMN} w^{PQR} \\
3(w^{L[JK} \delta^{I]}_M - \tfrac{1}{9} w^{IJK}\delta^L_M) &
\tfrac{1}{36\sqrt{3}} \epsilon^{ILKPQRLMN} v_{PQR} &
0
\end{matrix}
\right)
\non
\eena
and set $\V_- = \exp Z$. Finally, define $\V_+$ by
\bena
\V_+ = \left(
\begin{matrix}
V_I{}^A V^J{}_B & 0 & 0 \\
0 & V_{[I}{}^A V_J{}^B V_{K]}{}^C & 0 \\
0 & 0 & V^{[I}{}_A V^J{}_B V^{K]}{}_C
\end{matrix}
\right)
\eena
where $V_I{}^A$ and $V^I{}_A$ are inverses of each other. Then $\Phi$ is given by
\ben
\Phi = \V \cdot \tau(\V)^{-1} \ , \quad \V = \V_-\V_+ \ ,
\een
where $\tau$ is the symmetric space involution corresponding to the involution on the Lie algebra ${\frak e}_{8(+8)} \cong {\frak so}(16) \oplus {\frak h}$ which acts by $+1$ on the
first summand (the maximal abelian subalgebra) and by $-1$ on the second. With this definition of $\Phi$, the reduced action
again takes precisely the same form~\eqref{Sreduced}, up to a different, irrelevant, numerical prefactor in front of the scalar field term.

 One can now exploit divergence identities and prove, by the same general strategy as in the vacuum theory, that two non-extremal, single horizon, black hole solutions
whose corresponding interval structure $\{\underline{v}_J, l_J\}$, angular momenta $J_i$, and charges $Q_{\rm E}[C]$ all coincide, must in fact be the same.
To carry out these steps in detail, one must understand precisely how these data affect the boundary conditions
on the fields $\Phi_i, i=1,2$ of the two solutions under consideration. For this in turn, one must
understand how to describe all possible cycles $C \in H_7(\M, \mz)$ in terms of the data
$\{\underline{v}_J\}$, and how the information about the charges and angular momenta
translates into information about the boundary conditions of the fields $\Phi_i$
for the two solutions under consideration. This is rather complicated indeed, but it can be done~\cite{hollands12}.
Apart from the technical complications related to the structure of $E_{8(+8)}$,
the essential new feature compared to the vacuum theory are the charges
$Q_{\rm E}[C]$ defined relative to the various cycles $C$, which now enter the proof. This issue arises also--but can be understood much more easily--in the case Einstein-Maxwell theory in 5 dimensions, so we will now explain it in that theory.

\medskip
\noindent
{\bf $D=5$ Einstein-Maxwell theory:} As is seen from the above table, unlike minimal supergravity in $D=5$ dimensions, this theory actually does {\em not} have a sigma model formulation even in the presence of an isometry group $\mr \times U(1)^2$ of the spacetime. However, it {\em does} have a sigma model formulation if we make a additional,
by-hand, simplifying assumptions about the nature of the solutions to be considered. The simplifying assumptions are:
\begin{enumerate}
\item About the {\em spacetime metric} we assume that one of the axial Killing fields, say $\psi_1$, is orthogonal to the other Killing fields, $0=g(t,\psi_1)=g(\psi_1,\psi_2)$,
    and that it is hypersurface orthogonal, $\psi_1 \wedge \D \psi_1 = 0$.
\item About the {\em Maxwell field} we assume that there is a 1-form $\xi$ orthogonal
      to the Killing fields such that $F = \half \xi \wedge \psi_1$.
      It can easily be shown that, if the Maxwell field arises from a vector potential
      $F = \D A$ which is invariant under the Killing fields, then this will be the case if and only if $A$ is proportional to $\psi_1$ at each point in $M$.
      Note, however that we do {\em not} assume the existence of such a vector potential here.
\end{enumerate}

Let us first point out the main simplifications which follow from assumptions
1) and 2). The first immediate consequence of 1) is that $J_1=0$.
Secondly, because the Killing field $\psi_1$ is demanded to be orthogonal to $\psi_2$,
if $v_1 \psi_1 + v_2 \psi_2 = 0$ at a point in spacetime, then either $\underline{v} = (v_1, v_2) = (0,0)$,
or $\underline{v} = (0,1), (1,0)$, or both axial Killing fields vanish. Thus, the interval structure (see sec.~\ref{ref:wp}) of
any solution satisfying assumption 1) can only be of the following possibilities (i)---(iv):

\vspace{1cm}
\begin{center}
\begin{tabular}{|c|c|c|}
\hline
               & Moduli $l_J$ & Vectors $\underline{v}_J$ \\
\hline
(i) & $\infty, l_1, \dots, l_p, \infty $ & $(1,0), (0,1), \dots (1,0), (0,0), (1,0), (0,1)\dots, (0,1)$ \\
\hline
(ii) & $\infty, l_1, \dots, l_p, \infty $ & $(1,0), (0,1), \dots (0,1), (0,0), (0,1), (1,0) \dots, (0,1)$ \\
\hline
(iii) & $\infty, l_1, \dots, l_p, \infty $ & $(1,0), (0,1), \dots (0,1), (0,0), (1,0), (0,1) \dots, (0,1)$ \\
\hline
(iv) & $\infty, l_1, \dots, l_p, \infty $ & $(1,0), (0,1), \dots (1,0), (0,0), (0,1), (1,0) \dots, (0,1)$ \\
\hline
\end{tabular}
\end{center}

Thus, the possible interval structures are severely restricted by 1). By table~\ref{table1}, it then follows
that the only possible horizon topologies are
\begin{equation}
 B \cong S^1 \times S^2 \quad \text{(black ring)}, \quad B \cong S^3 \quad \text{(black hole),}
\end{equation}
with the first case realized when the vectors to the left and right of the horizon
$\underline{v}_{h-1}, \underline{v}_{h+1}$ are equal [i.e., for the interval structures (i) and (ii)] and the
second case realized when they are different [i.e., for the interval structures (iii) and (iv)].
In particular, the Lens-spaces $L(p,q)$ are excluded as possible horizon topologies by 1).

From 2), the electric charge $Q_{\rm E}[C] = 0$ associated with any 2-cycle $C$ vanishes, and the Maxwell field is completely characterized by the 1-form
\begin{equation}
f =  i_{\psi_1}F \, ,
\end{equation}
which is closed by the equations of motion for the Maxwell field, $\D f=0$. We define the twist 1-form by
\begin{eqnarray}
\omega = \half \psi_1 \wedge \psi_2 \wedge \D \psi_2 \, .
\end{eqnarray}
Again one can show that this is closed $\D \omega = 0$. It can then be shown that there exist
globally defined potentials $\D \alpha = f, \D \chi = \omega$.
If the Maxwell field arises from a globally defined vector potential, $F = \D A$---which we do {\em not} assume---then $\alpha = i_{\psi_1} A$.

\medskip
\noindent
Using the potentials $\alpha, \chi$, one can now write down the reduced Einstein-Maxwell equations on the orbit space $\hat M$, in the form of a pair of sigma model equations~\cite{hoya}:
Let $\nu, w, u$ be the functions on $\hat \M$  defined through:
\begin{equation}
e^{2u} = g(\psi_1, \psi_1) \, , \quad e^{-u+2w} = g(\psi_2, \psi_2) \, , \quad
e^{-u+2w+2\nu} = g(\D r, \D r) \, ,
\end{equation}
with $r$ defined as in eq.~\eqref{rdef}.
Then the complete Einstein-Maxwell equations are equivalent to the following pair of matrix
equations on the upper complex half plane $\hat \M=\{(r,z) \mid r>0\}$:
\begin{eqnarray}\label{smodel}
\D \star_2 \left(r \Phi^{-1}_{1} \D \Phi_{1}^{}\right)&=&0 \, ,\nonumber\\
\D \star_2 \left(r \Phi^{-1}_{2} \D \Phi_{2}^{}\right)&=&0 \, ,
\end{eqnarray}
together with an equation for $\nu$, which we do not write down.
The matrix fields are defined in terms of $u, w,\alpha,\chi$ by
\begin{eqnarray}
\Phi_{1} = \left(%
\begin{array}{cc}
  e^{u} + {1\over 3}e^{-u}\alpha^2 & {1\over \sqrt{3}} e^{-u}\alpha \\
 {1\over \sqrt{3}} e^{-u}\alpha & e^{-u} \\\end{array}%
\right) ,
\end{eqnarray}
and
\begin{eqnarray}
\Phi_{2} = \left(%
\begin{array}{cc}
  e^{2w} + 4\chi^2e^{-2w} & 2\chi e^{-2w} \\
 2\chi e^{-2w} & e^{-2w} \\\end{array}%
\right).
\end{eqnarray}
The first two equations state that
the matrix fields $\Phi_{1}$ and $\Phi_{2}$ each satisfy the equations of a 2-dimensional
sigma-model. The matrix fields are real, symmetric, positive definite, with determinant equal to $1$ on the interior of $\hat \M$, so we have as target space two decoupled copies of $SL(2)/SO(2)$. With the
aid of this sigma model formulation, one proves~\cite{hoya}:

\begin{thm}
 Consider two stationary, asymptotically flat, Einstein-Maxwell
black hole spacetime of dimension 5,  having  one
time-translation Killing field and two axial Killing fields.
We also assume that there are no points with discrete isotropy subgroup under the action of the
isometry group in the exterior of the black hole, and we
assume that the Killing and
Maxwell fields satisfy the assumptions 1) and 2) above. If
the two solutions have the same interval structures, the same
values of the mass $m$, same angular momentum  $J_2$, and same
magnetic charges $Q_{\rm M}[C_l] = (2\pi)^{-1} \int_{C_l} F$ for all 2-cycles $C_l \in H_2(\M, \mz)$, then they are isometric.
\end{thm}

The proof is very similar to that in the vacuum case, except for the following new consideration: To prove the boundedness of
the quantities analogous to $\sigma$ in~\eqref{sigmadef} built from the matrices $\Phi_1$ for
two given solutions as in the theorem, one also needs to use that
\begin{equation}
\alpha(z) - \alpha(z') = \frac{1}{2\pi} \int_C F = \frac{1}{2\pi} Q_{\rm M}[C] \, .
\end{equation}
where $z,z'$ are from two boundary intervals labeled by the same
integer vectors $(1,0)$ or $(0,1)$, for a suitable cycle $C \in H_2(\M)$. Knowing that the magnetic charges of both solutions are the same then enables one via this identity to show that boundary value $(r=0)$
of the $\alpha$'s for both solutions are also the same. An argument of this sort--but rather more complicated--also
works in 11-dimensional supergravity.

The proof shows that the non-trivial 2-cycles [i.e., basis elements of $H_2(\M)$] in the exterior of the spacetime may in fact be obtained as follows. We know that the real axis $\{r=0\}$ bounding $\hat \M=\{(r,z) \mid r>0\}$ is divided into intervals, each labeled with an
integer 2-vector $\underline{v}_J = (1,0)$ or $\underline{v}_J = (0,1)$, see the above table.
Now consider all possible curves $\hat \gamma_p, p=1,2,\dots$ in $\hat \M$ with the property that $\hat \gamma_p$ starts at $z$
on an interval labeled $(1,0)$, and ends at $z'$ on another interval labeled $(1,0)$, with no interval with label $(1,0)$ in
between. If we now lift $\hat \gamma_p$ to a curve $\gamma_p$ in $\M$, and act with all isometries generated by
$\psi_1$ on the image of this curve,
then we generate a closed 2-surface $C_p$ in $\M$
which is topologically a 2-sphere for all $p$, see the following picture.

\begin{center}
\begin{tikzpicture}[scale=1.1, transform shape]
\shade[left color=gray] (-4,0) -- (5,0) -- (5,3)  -- (-4,3) -- (-4,0);
\draw[very thick,color=red] (1,0) -- (3,0);
\draw[very thick,color=red,dashed] (3,0) -- (4,0);
\draw[->,very thick,color=red] (4,0) -- (5,0);
\draw[very thick,color=red] (-4,0) -- (0,0);
\draw[very thick,color=blue] (0,0) -- (1,0);
\draw (.5,0) node[below,color=blue]{$\hat B = B/U(1)^2$};
\draw (3.6,2.8) node[below]{$\hat \M=\{(r,z) \mid r>0\}$};
\draw (3,0) node[below]{$z'$};
\draw (-2.8,0) node[below]{$z$};
\draw[black,thick] (3,0) -- (3,2.1) -- (-2.8,2.1) -- (-2.8,0);
\draw[black,thick] (4.5,0) -- (4.5,1.4) -- (-2.8,1.4) -- (-2.8,0);
\draw (-0.2,1.4) node[below]{$\hat \gamma_{2}=C_2/U(1)^2$};
\draw[black] (4.5,0) node[below]{$z''$};
\draw (-0.2,2.1) node[above]{$\hat \gamma_1=C_1/U(1)^2$};
\filldraw[color=blue] (1,0) circle (.05cm);
\filldraw[color=blue] (0,0) circle (.05cm);
\filldraw[color=red] (-1.8,0) circle (.05cm);
\draw[color=red] (-1.8,0) node[below]{$z_{J+1}$};
\filldraw[color=red] (-3.7,0) circle (.05cm);
\draw[color=red] (-3.7,0) node[below]{$z_{J}$};
\filldraw[color=red] (-1,0) circle (.05cm);
\filldraw[black] (4.5,0) circle (.05cm);
\filldraw[color=red] (2,0) circle (.05cm);
\filldraw[color=red] (2.5,0) circle (.05cm);
\filldraw[color=black] (-2.8,0) circle (.05cm);
\filldraw[color=black] (3,0) circle (.05cm);
\end{tikzpicture}
\end{center}

We may repeat this by replacing $\hat \gamma_p, p=1,2, \dots$ with a set of curves each
starting on an interval labeled $(0,1)$, and ending on another interval labeled $(0,1)$, with no interval with label $(0,1)$ in between. If we again lift these curves to curves in $\M$, and act with all isometries generated by
$\psi_2$, then we generate a set of topologically inequivalent closed 2-surfaces $\tilde C_q, q=1,2, \dots$ in $\M$, each of which is topologically a 2-sphere. It may be seen that the set of 2-surfaces $\{ C_p, \tilde C_q \}$ forms a basis of $H_2(\M)$, and also of $H_2(\Sigma)$, where the 4-manifold $\Sigma$ is a spatial slice
going from infinity to the horizon (so that topologically $\M=\mr \times \Sigma$). In this 4-manifold, we can compute
intersection numbers as $C_p: \tilde C_q = \pm 1$ or $=0$, depending on whether the corresponding
curves in $\hat \M$ intersect or not. The rank of $H_2(\Sigma)=H_2(\M)$ in the cases (i) through (iv) in the above table, and the intersection matrix $Q_\Sigma: H_2(\Sigma) \times H_2(\Sigma) \to \mz$ is therefore easily computed to be given by
\ben
Q_\Sigma =
\bigoplus^m \left(
\begin{matrix}
0 & 1 \\
1 & 0
\end{matrix}
\right)
\een
where $m$ is related to the number of intervals in the interval structure (compare eq.~\ref{qform}).
Only the magnetic charges $Q_{\rm M}[C_p]$ enter in the proof of the above theorem. The magnetic charges
$Q_{\rm M}[\tilde C_q]$ are not needed and in fact vanish, due to assumptions 1) and 2) at the beginning of this
section. Thus, for the simplest interval structure $(0,1),(0,0),(1,0)$, there are no non-trivial magnetic
charges, and the unique solution within the class studied here
is completely specified by $J_2, m$. In fact, this unique solution is
the Myers-Perry black hole~\cite{MP86}, with vanishing Maxwell field.

\subsection{Static black holes}

To keep the discussion simple, we have considered above uniqueness theorems for non-extremal static black holes
 only within vacuum Einstein gravity, and for asymptotically flat boundary conditions. There are also uniqueness theorems for other theories, and other boundary conditions (although not for KK-type asymptotic conditions in as far as we are aware), as well as for extremal black holes in Einstein-Maxwell theory.
 We summarize some results in this direction.

\begin{enumerate}
\item {\em Dilatonic black holes}: In higher dimensions, we have the exact
solutions of dilatonic Einstein-Maxwell theory \cite{Gibbons:1987ps}.
One can prove their uniqueness in the same manner
here \cite{GIS2}. The same technique also applies to the case
of Einstein coupled to sigma-model fields~\cite{Rogatko02}.

\item
{\em Einstein-form fields}: A uniqueness theorem Einstein-$p$-form fields
when $p\geqslant (D+1)/2$ was obtained in~\cite{EOS}.
In the proof of this theorem, one uses the same kind of conformal transformation
employed in the proof of the Schwarzschild-Tangherlini spacetime above in sec.~\ref{sec:static}.

\item
{\em Extremal black holes in Einstein-Maxwell theory}:
 As already mentioned, in the extremal limit, there are exact solutions
for charged multi-black holes. The uniqueness of such solutions has been proven
\cite{CRT06,Rogatko03}. See also~\cite{Rogatko05} for
related work.

\item
{\em  Asymptotically de Sitter/anti-de Sitter}:
There are exact static solutions of Einstein-gravity $+$ cosmological constant,  the
Schwarzschild-(anti-)de Sitter solutions.
In~\cite{Boucher:1983cv,ACD02} attempts have been made to generalize
the static uniqueness to the asymptotically anti-de Sitter case, but the situation remains open. In particular, recent numerical investigations~\cite{bizon}
indicate that AdS-spacetime is unstable to perturbations, and the end state
of the dynamical evolution of an instability may well be an unknown, non-standard black hole
spacetime. See also Ref.~\cite{Kodama.H2004} for perturbative investigations of potential
new families of static black holes in these theories.
\end{enumerate}

\subsection{Supersymmetric black holes}
\label{sec:susy}
An entirely different approach to the classification of black hole spacetimes,
applicable to supergravity theories, is to consider solutions with Killing {\em spinors},
rather than Killing {\em vectors}. Unlike the Killing vector condition, $\pounds_t g = 0$,
which takes the same form in any theory, the precise form of the Killing spinor condition depends on the theory under consideration. In 4 dimensional $N=2$ supergravity,
this analysis has been completed first by~\cite{tod}, but the method is applicable, in principle, in any dimension in which supergravity theories can exist (i.e. $D\le 11$). Consider for example the case of minimal supergravity in
$D=5$ dimensions, with action\footnote{In this section, the signature is $(+----)$, opposite
from the rest of the paper.}
\ben\label{5dsusy}
S = \int_\M \frac{1}{2} R \star_5 1 + F \wedge \star_5 F + \frac{4}{3\sqrt{3}} F \wedge F \wedge A \ ,
\een
where $A$ is a one-form with field strength $F = \D A$. The above action must be supplemented in the full supergravity theory
by terms involving fermionic partner fields with half integral spin. But at the classical level, it does not make much physical sense to consider solutions in which the fermionic fields are non-trivial, and one thus sets them to zero. The full action with
superpartners is invariant under the odd supersymmetry transformation $\delta_\epsilon$ depending linearly on a
symplectic Majorana spinor field\footnote{This is a pair of $4$-dimensional complex spinor fields
$\epsilon^a$, $a=1,2$, subject to the condition $\bar \epsilon^a = (\epsilon^a)^T C$,
with $C$ the charge conjugation endomorphism of the Clifford algebra defined by $C (\gamma_\mu)^T C^{-1} = \gamma_\mu$. The conjugate spinor is
defined by $\bar \epsilon^a \ \delta_{ab} = (\epsilon^a)^* \beta \ \epsilon_{ab}$, where
$\beta$ is the endomorphism of the Clifford algebra defined by $\beta (\gamma_\mu)^* \beta^{-1} = \gamma_\mu$,
and where $*$ is hermitian adjoint.}
$\epsilon^a$, $a=1,2$, which maps the bosonic fields $g,A$ to a combination of
fields depending on the fermionic superpartners, and which maps the fermionic superpartners to an
expression involving the bosonic fields. Let us demand that a purely bosonic field configuration be ``supersymmetric'',
in the sense that it is annihilated by $\delta_\epsilon$. Since the fermionic fields vanish by assumption, we automatically have $\delta_\epsilon g = 0 = \delta_\epsilon A$, whereas the condition that $\delta_\epsilon$ on the fermionic fields (spin 3/2-partner of the metric)
gives zero amounts to\footnote{$\gamma_{\mu_1...\mu_n} = \gamma_{[\mu_1} \cdots \gamma_{\mu_n]}$ denote the
generators of the Clifford algebra ${\rm Cliff}(T\M)$.}
\ben
\left(\nabla_\mu + \frac{1}{4\sqrt{3}}(\gamma_{\mu\nu\sigma} + g_{\mu\nu} \gamma_\sigma ) F^{\nu\sigma}
\right) \epsilon^a = 0 \ .
\een
This is the Killing-spinor condition for $D=5$ minimal supergravity. The existence of a Killing spinor is very restrictive--much more so than
demanding merely the existence of a Killing vector. This can be seen e.g. from the fact that, since a supersymmetry transformation ``squares to an infinitesimal translation'' (i.e. a Lie-derivative), the ``square'' of a Killing-spinor
automatically has to be a Killing vector, and also a symmetry of the other bosonic fields in the theory, i.e.
$F$ in our case. More precisely, define
\ben
t^\mu  := \half \epsilon_{ab} \ \bar \epsilon^a \gamma^\mu \epsilon^b \ ,
\een
where $\epsilon^{ab}$ is the standard 2-dimensional symplectic matrix. Then $t$ is necessarily a
time-like, or null, Killing vector field, $\pounds_t g = 0$,
and one also necessarily has $\pounds_t F = 0$. If $(\M, g, A)$ represents an asymptotically flat black hole spacetime, then $t$ is by construction tangent to the generators of the horizon, and because it is timelike or null both inside
and outside the horizon, and null on the horizon, we must have $\D( g(t,t) ) = 0$ on the horizon. Using the standard
formula for the surface gravity $\D( g(t,t) ) = -2\kappa \ t$ then
shows that $\kappa = 0$, i.e. the black hole is necessarily extremal. In fact, the existence of a Killing spinor
is even more stringent than the above argument suggests, because it implies many more differential relations
than just the Killing vector equation. These have been exploited systematically by~\cite{harveygauntlett} (and
previously by Tod~\cite{tod} in the case of $N=2$ supergravity in $D=4$). \cite{harveygauntlett} proceed by defining the
real tensorial quantities $f, X_1, X_2, X_3$ by
\ben
f \ \epsilon^{ab} = \bar \epsilon^a \epsilon^b \ , \quad
\left(
\begin{matrix}
X_1 + iX_2 & iX_3\\
-iX_3 & X_1 - iX_2
\end{matrix}
\right)^{ab} = \bar \epsilon^a \gamma_{\mu\nu} \epsilon^b \ \D x^\mu \wedge \D x^\nu \ .
\een
The field equations and Killing spinor equation then imply, among other things, the following further relations:
\bena
f^2 = g(t,t) \ , \quad
\D f = -\frac{2}{\sqrt{3}} i_t F \ , \quad
\D X_i = 0 \ , \quad
\D \star_5 X_i = -\frac{2}{\sqrt{3}} F \wedge X_i \ .
\eena
It turns out that, starting from these equations, one can
locally determine~\cite{harveygauntlett} all possible analytic field configurations with Killing spinor, which can then be analytically continued. In this way, one can obtain, in principle, a classification (which
includes black holes, but also other types of spacetimes) of supersymmetric bosonic configurations $(g,A)$. We think that it would be worthwhile to understand in detail the
global structure of these solutions. Progress towards this goal has been made in the special case of spherical horizons by~\cite{reallsusy}.

A similar analysis can presumably be carried out in other, more complicated, supergravity theories. Of particular interest would be a similarly complete classification of
solutions with various numbers of independent Killing spinors in 11-dimensional supergravity.
Progress in this direction has been made e.g. in the papers~\cite{gauntlet1,gauntlet2},
see also references therein.

Thus, in supersymmetric theories, the classification of black hole solutions via Killing-spinors seems a feasible--albeit complicated--task. However, one cannot hope to obtain in this way solutions at non-zero temperature, i.e. non-extremal ones.

\section{Summary and open issues}

In this review, we have tried to give an overview about what is known about black hole uniqueness theorems for
higher dimensional, stationary, black hole spacetimes. Along the way, we have described several related general structural results
 about such black holes such as the topology, rigidity, and staticity theorems, which are also of independent interest. These arguments involve a considerable breadth of mathematical methods, from
differential geometry, to topology, to group theory, to ergodic theory, and to PDE theory, which are combined in
a non-trivial fashion. As we have indicated, the state of knowledge concerning higher dimensional
black holes in general, and uniqueness theorems in particular, is much less satisfactory than in four dimensions,
although also in four dimensions there are still many open issues for many theories other than Einstein-Maxwell theory. In general, it seems that our knowledge about static or supersymmetric black holes in higher dimensions is more advanced
than for stationary solutions, but there are also still important open issues for static and supersymmetric black holes.
As open issues that in our opinion deserve attention we would like to mention:

\begin{enumerate}
\item To obtain uniqueness theorems for stationary higher dimensional black holes, one needs to assume
more symmetries than seem to be generic on the basis of the investigations~\cite{blackfold,Dias}, and one is at the moment restricted
to models in which the dimensionally reduced theories can be cast into the framework of sigma-models
with negatively curved target spaces. There are some ideas how to replace the interval structure
that seems important in existing theorems~\cite{Harmark09,HHI10}, but more ideas are needed.

\item To obtain uniqueness theorems for static black holes, one is for the moment restricted to
asymptotically flat boundary conditions, excluding thus e.g. the, very interesting and relevant, Kaluza-Klein type
boundary conditions.

\item It would also be important to achieve a full classicfication of supersymmetric black holes in more complicated supergravity theories than have been analyzed so far, in particular in 11-dimensional supergravity.
\end{enumerate}

We feel that input from numerical methods will be needed to get a more complete picture of the landscape of
higher dimensional black holes, and to help answer some of these questions in particular.

\section*{Acknowledgments}
This work is supported in part (AI) by the Grant-in-Aid for Scientific Research Fund of the JSPS (C)No. 22540299. We would like to thank the referee for his careful reading of this manuscript.

\end{document}